\documentclass[fleqn,usenatbib]{mnras}

\usepackage{newtxtext,newtxmath}

\usepackage[T1]{fontenc}
\usepackage[normalem]{ulem} 
\DeclareRobustCommand{\VAN}[3]{#2}
\let\VANthebibliography\thebibliography
\def\thebibliography{\DeclareRobustCommand{\VAN}[3]{##3}\VANthebibliography}

\usepackage{graphicx}	
\usepackage{amsmath}	
\usepackage{amsfonts}
\usepackage{threeparttable}

\newcommand{\msun}{\mbox{M$_{\odot}$}}

\newcommand{\gcc}{g~cm$^{-3}$}

\newcommand{\edot}{$\dot{E}$}
\newcommand{\fss}{\hbox{$.\!\!^\mathrm{s}$}}        
\newcommand{\h}{$^\mathrm{h}$}
\newcommand{\m}{$^\mathrm{m}$}

\newcommand{\gr}{$\gamma$-ray}
\newcommand{\psr}{J0554}
\newcommand{\jpsr}{J0554+3107}
\newcommand{\snr}{G179.0+2.6}
\newcommand{\ergs}{erg~s$^{-1}$}

\newcommand{\phs}{ph~cm$^{-2}$~s$^{-1}$~keV$^{-1}$}
\newcommand{\xmm}{\textit{XMM-Newton}}

\newcommand{\fermi}{\textit{Fermi}}
\newcommand{\sw}{\textit{Swift}}
\newcommand{\degs}{\ifmmode ^{\circ}\else$^{\circ}$\fi}
\newcommand{\amin}{\ifmmode ^{\prime}\else$^{\prime}$\fi}
\newcommand{\asec}{\ifmmode ^{\prime\prime}\else$^{\prime\prime}$\fi}
\newcommand{\tc}{$t_\mathrm{c}$}

\newcommand{\nh}{$N_\mathrm{H}$}
\newcommand{\chir}{$\chi^2_\nu$}
\newcommand{\nsl}{$(\mathrm{NS130 + PL}) \times \textsc{gabs}$}
\newcommand{\nsh}{$(\mathrm{NS133 + PL}) \times \textsc{gabs}$}
\newcommand{\nsm}{$(\textsc{nsmdintb} + \mathrm{PL}) \times \textsc{gabs}$}
\newcommand{\bbg}{(2BB + PL) $\times$ {\textsc{gabs}}}
\newcommand{\dof}{d.o.f.}

\title[\xmm\ observations of PSR J0554+3107]{\xmm\ observations of PSR J0554+3107: 
pulsing thermal emission from a cooling high-mass neutron star}

\author[A. S. Tanashkin et al.]{
A. S. Tanashkin,\thanks{E-mail: artyom.tanashkin@gmail.com}
A. V. Karpova,
A. Y. Potekhin,
Y. A. Shibanov and
D. A. Zyuzin
\\
Ioffe Institute, Politekhnicheskaya 26, St. Petersburg, 194021, Russia
}

\date{Accepted XXX. Received YYY; in original form ZZZ}

\pubyear{2022}

\begin{document}
\label{firstpage}
\pagerange{\pageref{firstpage}--\pageref{lastpage}}
\maketitle

\begin{abstract}
\xmm\ observations of the middle-aged 
radio-quiet \gr\ pulsar \jpsr\ allowed us, for the first time, 
firmly identify it in X-rays by detection of pulsations with the pulsar period. 
In the 0.2--2~keV  band, the pulse profile shows 
two peaks separated by about a half of the rotation phase with the pulsed fraction 
of $25\pm6$ per cent. The  profile and spectrum 
in this band can be mainly described by thermal emission from the neutron star with 
the hydrogen atmosphere, dipole magnetic field of $\sim 10^{13}$~G 
and non-uniform surface temperature. 
Non-thermal emission from the pulsar magnetosphere is marginally detected 
at higher photon energies. The spectral fit with the atmosphere$+$power law 
model implies that \jpsr\ is a rather heavy and cool neutron star with the mass of 
1.6--2.1~\msun, the radius of $\approx 13$~km and the redshifted effective 
temperature of $\approx 50$~eV. 
The spectrum shows an absorption line of unknown nature at $\approx 350$~eV.
Given the extinction--distance relation, the pulsar is located 
at $\approx 2$~kpc and has the redshifted bolometric thermal luminosity of 
$\approx 2\times10^{32}$~\ergs. 
We discuss cooling scenarios for J0554+3107 considering 
plausible equations of state of super-dense matter inside 
the star, different  compositions of the heat-blanketing 
envelope and various ages. 
\end{abstract}

\begin{keywords}
stars: neutron -- pulsars: general -- pulsars: individual: PSR \jpsr
\end{keywords}



\section{Introduction}
\label{sec:intro}

Studies of thermal emission from neutron stars (NSs) is one 
of the ways to investigate the properties of super-dense nuclear 
matter in their interiors \citep[e.g.][]{yakovlev&pethick2004}.
If thermal emission originates from the entire NS surface, 
the comparison of the measured thermal luminosity (or, equivalently, mean effective temperature) 
with predictions 
of NS cooling theories can set constraints on the equation of state (EoS)
of such matter. Middle-aged (10$^4$--10$^6$~year-old) stars are
the best targets for such analysis since thermal components usually 
dominate over non-thermal ones in their X-ray spectra. 
However, effective temperatures and ages are estimated only for
a few dozens of NSs, which is not enough to make definite 
conclusions \citep{potekhin2020}. 

\begin{table}
\renewcommand{\arraystretch}{1.2}
\caption{\psr\ parameters obtained from \citet{pletsch2013}. 
Numbers in parentheses denote 1$\sigma$ uncertainties relating to the last 
significant digit quoted.}
\label{tab:pars}
\begin{center}
\begin{tabular}{lc}
\hline
R.A. (J2000)                                        & 05\h54\m05\fs01(3) \\
Dec. (J2000)                                        & +31\degs07\amin41\asec(4) \\
Galactic longitude $l$, deg                         & 179.058 \\
Galactic latitude $b$, deg                          & 2.697 \\
Spin period $P$, ms                                 & 465      \\
Spin frequency $f$, Hz                              & 2.15071817570(7)       \\
Frequency derivative $\dot{f}$, Hz s$^{-1}$         & $-$0.659622(5) $\times$ 10$^{-12}$ \\
Frequency second derivative $\ddot{f}$, Hz s$^{-2}$ & 0.18(2) $\times$ 10$^{-23}$ \\
Epoch of frequency determination, MJD               & 55214 \\
Data time span, MJD                                 & 54702--56383 \\
Solar system ephemeris model                        & DE405 \\
\hline
Characteristic age \tc, kyr                         & 51.7 \\
Spin-down luminosity \edot, \ergs\                  & 5.6 $\times$ 10$^{34}$ \\
Characteristic magnetic field $B_\mathrm{c}$, G     & 8.2 $\times$ 10$^{12}$ \\
\hline
\end{tabular}
\end{center}
\end{table}

The middle-aged radio-quiet PSR \jpsr\ (hereafter \psr) was discovered by 
Einstein@Home\footnote{\url{http://einstein.phys.uwm.edu}} 
in a blind search of \fermi\ Large Area Telescope (LAT) 
\gr\ data \citep{pletsch2013}. Its parameters are presented in
Table~\ref{tab:pars}. Upper limits on the integral flux density at 
111, 150 and 1400 MHz are 0.5, 1.2 and 0.066~mJy, respectively
\citep{tyulbashev2021,griesmeier2021,pletsch2013}.

The pulsar is projected onto the supernova remnant (SNR) \snr\ 
and likely associated with it \citep{pletsch2013}.
This is a shell-type oxygen-rich remnant \citep{how2018} 
with the diameter of about 70~arcmin. Its age is unclear: 
the large diameter and the low surface brightness imply the age 
of $\sim$~10--100~kyr, but the radial configuration of the magnetic field
is typical for young SNRs \citep{fuerst&reich1986,gao2011}.
There are different estimates of the distance
to the remnant, from about 3 to 6~kpc \citep{case&bhattacharya1998,guseinov2003,pavlovic2014}, 
obtained through empirical correlations between the radio surface brightness
and the diameter of the SNR.
However, recent studies show that \snr\ can be much closer, 
at $\approx 0.9$~kpc \citep{zhao2020}. This result is based on an apparent interstellar extinction jump 
at this distance along the remnant line of sight and on the assumption that it is associated with 
the dust formed by the SNR. However, the detected extinction jump 
may correspond to a foreground molecular cloud and not the remnant itself. 
If so, this value should be considered as the lower limit to the distance. 

Since \psr\ has not been detected in radio, it is not possible to obtain the distance basing 
on the dispersion measure. The only available estimate is
the so-called `pseudo-distance' of $\approx 1.9$~kpc 
calculated from the empirical relation between the distance and the pulsar
\gr\ flux \citep{sazparkinson2010}. 
Though such an estimate is 
rather uncertain (within a factor of 2--3), it is consistent with the association of \psr\ with \snr. 

\citet*{zyuzin2018} found the likely X-ray counterpart of \psr\ 
in the 1st \sw-XRT Point Source Catalogue  
\citep[1SXPS][]{evans2014}. Seventeen counts were detected from the source
in the 11~ks exposure, and thirteen of them are in the 0.3--1~keV band, 
which indicates that the source is soft. Indeed, fitting 
the X-ray spectrum with a power law model,
\citet{zyuzin2018} obtained the photon index $\Gamma>4$,
which is too high for pulsars
and may indicate the presence of a thermal-like spectral component. Pure thermal emission models~-- the black body  
and the NS atmosphere~-- resulted in NS surface temperatures of $\approx$~50--100~eV. 
In any case, the obtained model parameters remained very uncertain because of the low \sw\ count statistics. 
We therefore performed deeper observations with \xmm\ to confirm the counterpart of \psr\ 
and clarify its X-ray properties.

Here we present results of these observations. The data and the reduction procedure are presented 
in Sec.~\ref{sec:data}. Imaging is described in Sec.~\ref{sec:imaging}. Timing and spectral analyses 
are presented correspondingly in Sec.~\ref{sec:timing} and Sec.~\ref{sec:spec}. 
We discuss the results in Sec.~\ref{sec:discussion} and make conclusions
in Sec.~\ref{sec:sum}.


\section{Observations and data reduction}
\label{sec:data}

\begin{figure}
\begin{minipage}[h]{1.\linewidth}
\center{\includegraphics[width=1.0\linewidth,clip]{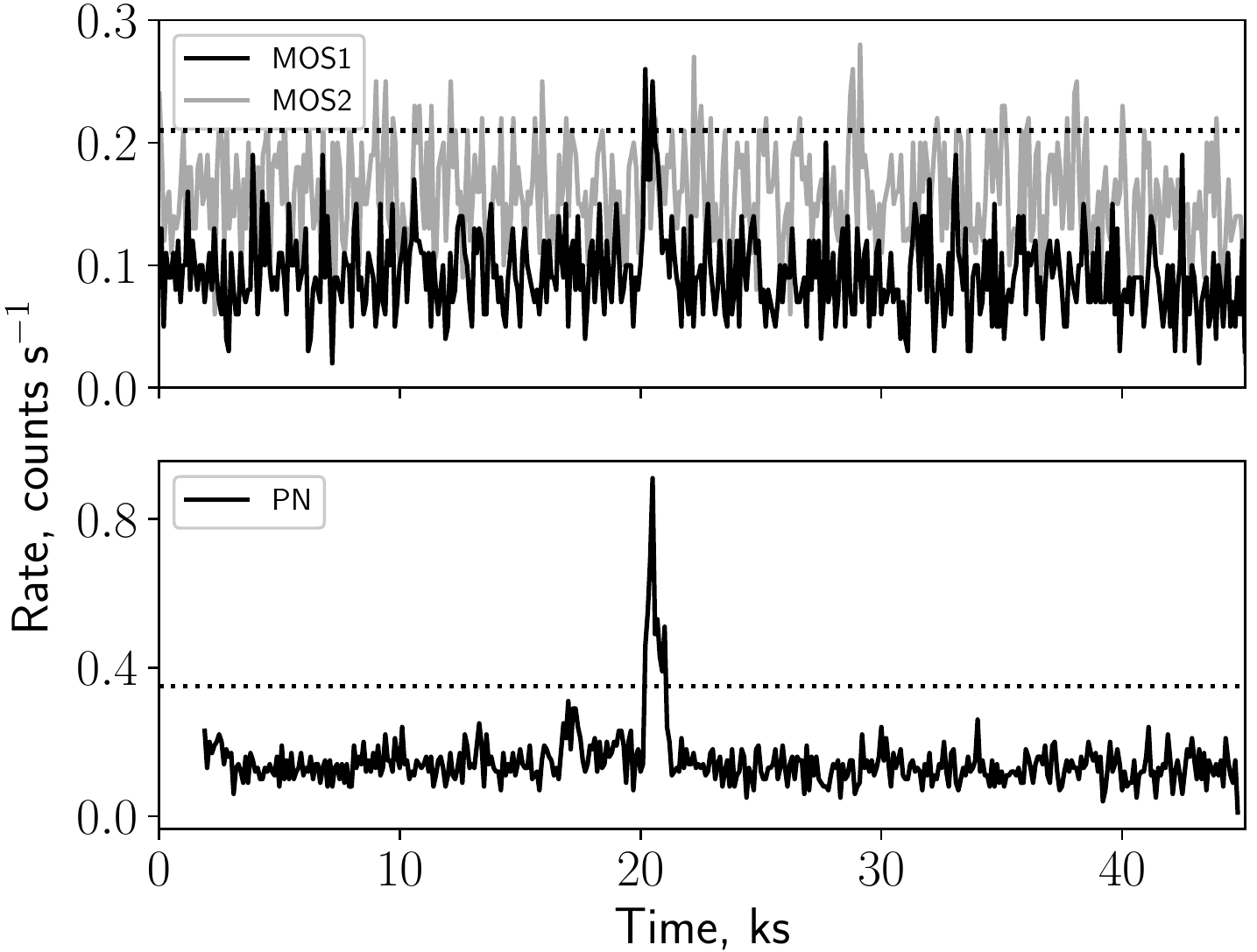}}
\end{minipage}
\caption{High-energy light curves extracted from the FoVs of 
the MOS ($>10$~keV; \textit{top}) and PN (10--12~keV; \textit{bottom}) cameras. 
The dotted lines show the thresholds applied to filter the MOS1 and PN data 
for periods of background flaring activity (no filtering was applied to the MOS2 data).}
\label{fig:lc}
\end{figure}

45-ks \xmm\ observations of the \psr\ field were carried out 
on 2021 October 7 (ObsID 0883760101, PI A. Karpova). 
The European Photon Imaging Camera Metal Oxide Semiconductor 
(EPIC-MOS) detectors were operated in the full frame mode 
and the EPIC-pn (PN hereafter) detector~-- in the large window mode.
The corresponding imaging areas are 28~arcmin~$\times$ 28~arcmin 
and 13.5~arcmin~$\times$ 26~arcmin. For all instruments, the thin filter 
was chosen.

We used the \xmm\ Science Analysis Software ({\sc xmm-sas}) v.19.1.0
for analysis. The {\sc emproc} and {\sc epproc} routines were utilised
to reprocess the data. To filter out the periods of background flares, 
we extracted high-energy light curves from the fields of view (FoVs) 
of all EPIC detectors. Only one short flare is present
(see Fig.~\ref{fig:lc}), mostly affecting the PN detector, but also 
having a slight impact on the MOS1 light curve, so we cleaned only 
the PN and MOS1 data, leaving the MOS2 data unfiltered.
As a result, the effective exposures 
are 44.3, 44.5 and 38.8~ks for MOS1, MOS2 and PN cameras, respectively. 
Single, double, triple and quadruple pixel events were selected for MOS 
({\sc pattern}~$\leq$ 12) and single and double pixel events~-- for PN 
({\sc pattern}~$\leq$ 4).

\section{Imaging}
\label{sec:imaging}

\begin{figure}
\begin{minipage}[h]{1.\linewidth}
\center{\includegraphics[width=1.0\linewidth,clip]{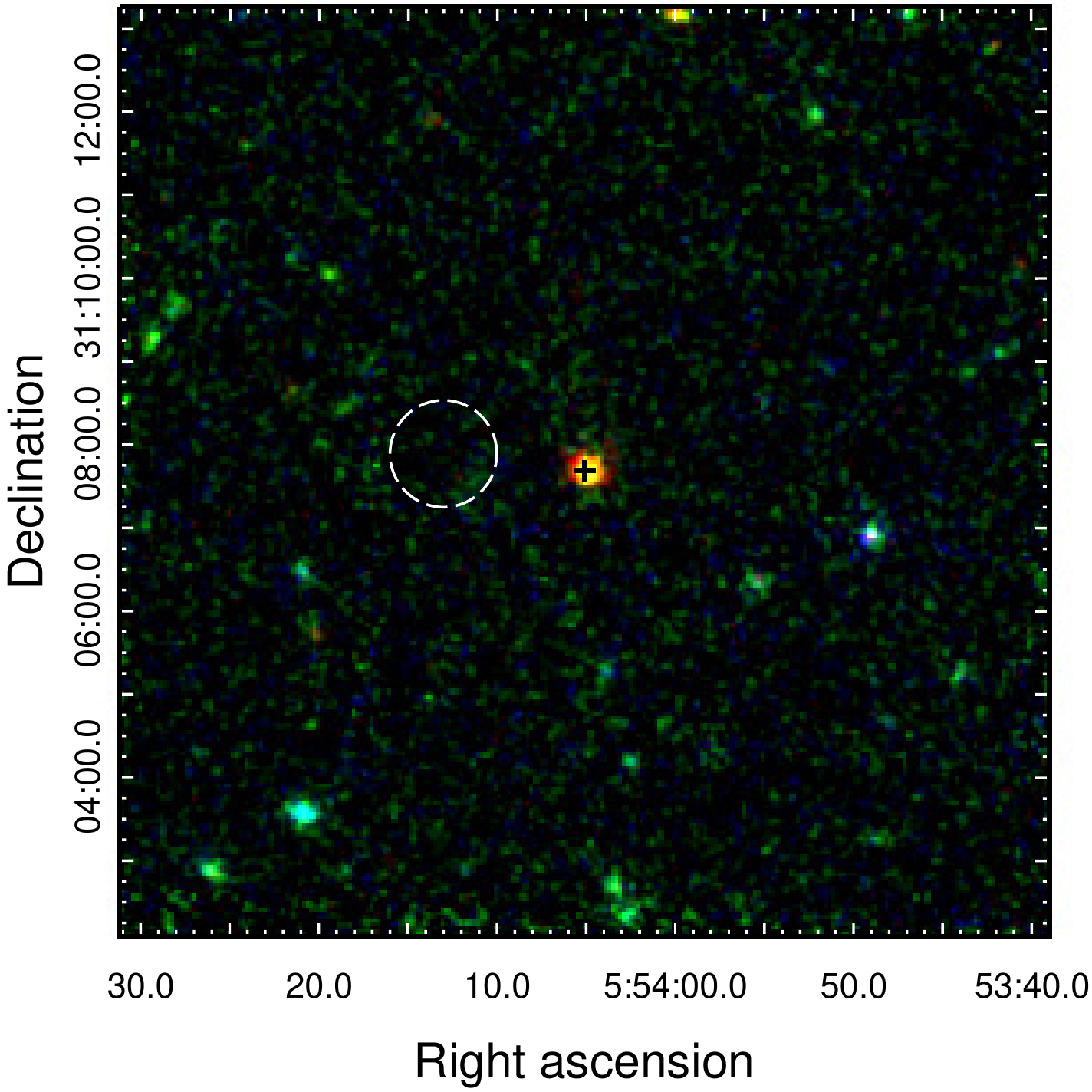}}
\end{minipage}
\caption{\xmm\ combined (MOS1+MOS2+PN) image of the \psr\ field obtained in soft (0.2--1~keV, red), 
medium (1--2~keV, green) and hard  (2--12~keV, blue)  energy bands. 
The bright soft point source in the centre of the image is the X-ray 
counterpart of the pulsar, whose $\gamma$-ray position is marked by the cross. The position uncertainty is
significantly smaller than the marker size and is not shown.
The dashed 38.5-arcsec radius circle shows the region used for the background extraction in the timing and spectral analyses.
}
\label{fig:img}
\end{figure}

\begin{figure}
\begin{minipage}[h]{1.\linewidth}
\center{\includegraphics[width=1.0\linewidth,clip]{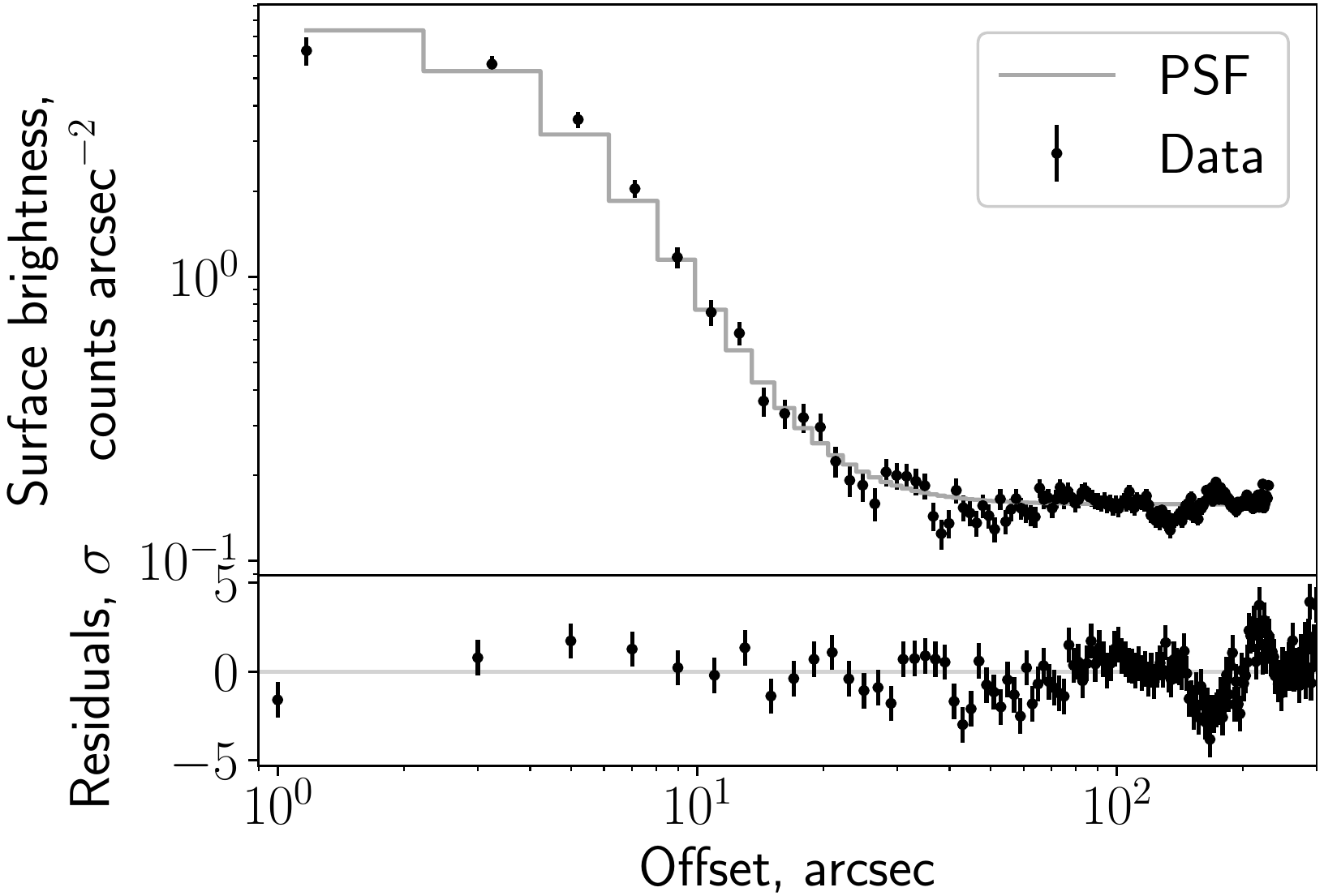}}
\end{minipage}
\caption{Comparison of the \xmm\ PSF and the observed radial brightness profile
obtained from the PN data in the 0.2--10~keV energy band.}
\label{fig:radprof}
\end{figure}

Using the `images' 
script\footnote{\url{https://www.cosmos.esa.int/web/xmm-newton/images}} 
\citep{imagesscript}, we created combined MOS+PN exposure-corrected
images in different energy bands. The resulting image is shown in 
Fig.~\ref{fig:img}. The \psr\ candidate counterpart is seen as a bright source in the centre.
Its position 
R.A.~= 05\h54\m05\fss067(10) and Dec.~= +31\degs07\amin41\farcs40(13)
was derived by the {\sc edetect\_chain} task using the data from all EPIC
detectors (numbers in parentheses are 1$\sigma$ pure statistical
uncertainties). Taking into account the \xmm\ absolute pointing accuracy of 
1.2~arcsec\footnote{\label{CAL-TN-0018}\url{https://xmmweb.esac.esa.int/docs/documents/CAL-TN-0018.pdf}}, 
the coordinates are in agreement within 1$\sigma$ with the \psr\ ones
obtained from the \fermi\ data (Table~\ref{tab:pars}).

From Fig.~\ref{fig:img} one can see that the \psr\ candidate counterpart is 
a soft source. No extended emission like a pulsar wind nebula (PWN) or
misaligned outflows are seen in the pulsar vicinity. However, the \xmm\
point spread function (PSF) is rather wide and the compact PWN if exists can 
be blurred with the pulsar. 
To treat this possibility more carefully we used the {\sc eradial} 
tool, which extracts the source radial brightness profile and fits the PSF to it. The result is shown 
in Fig.~\ref{fig:radprof}, and we can conclude that
the radial profile is consistent with the PSF and no compact PWN is resolved. 
We also do not resolve any extended emission in the \xmm\ FoV  
that could be identified with the SNR \snr, whose radio shell is outside
the FoV.


\section{Timing analysis}
\label{sec:timing} 

\begin{figure}
\begin{minipage}[h]{1.\linewidth}
\center{\includegraphics[width=1.0\linewidth,clip]{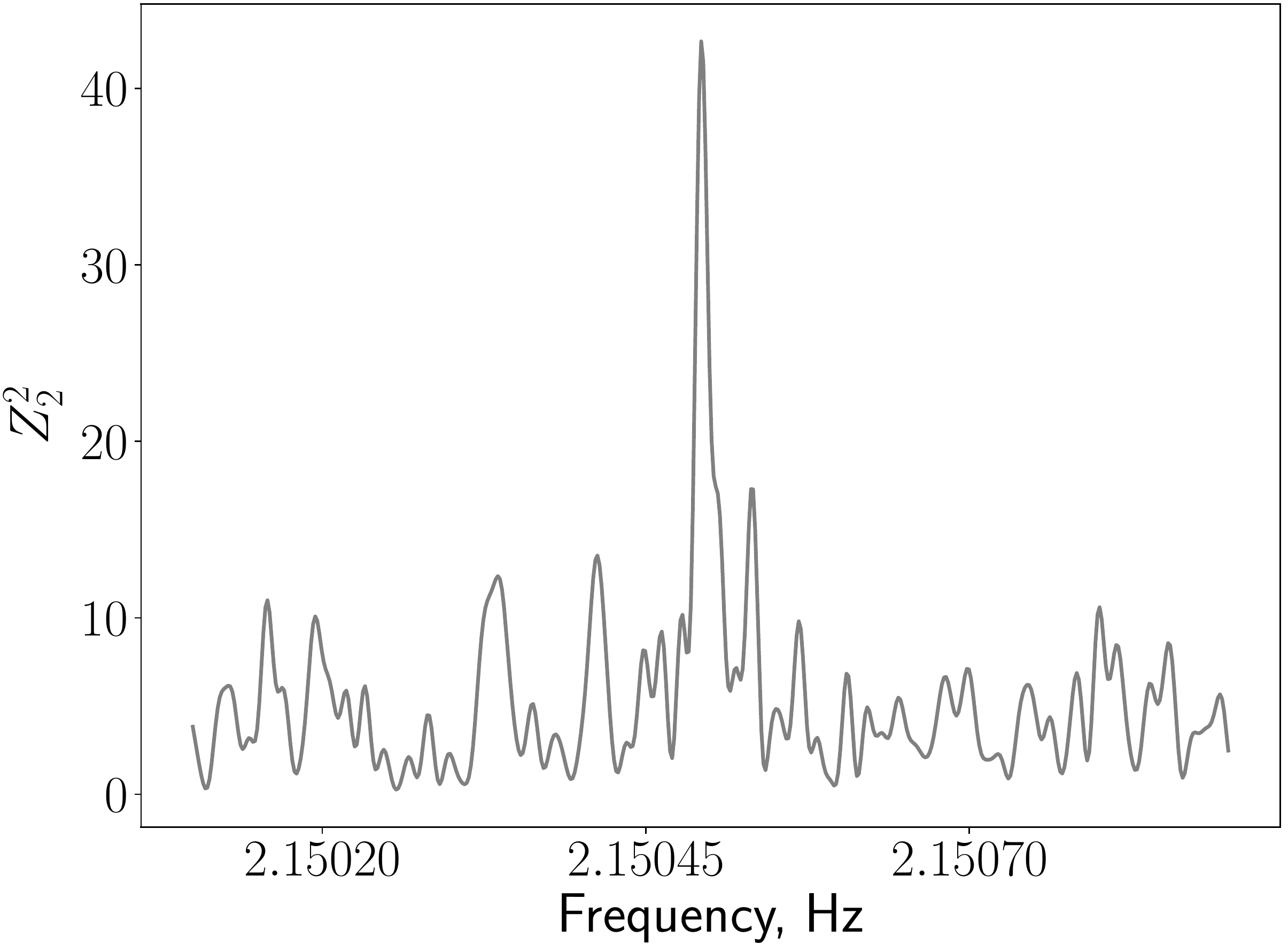}}
\end{minipage}
\caption{$Z^2_2$-test periodogram for \psr\ in the 0.2--2~keV energy band.
}
\label{fig:ztest}
\end{figure}

The PN detector operating in the large window mode provides 
the time resolution of $\approx 48$~ms, which is sufficient to search for regular X-ray pulsations 
with the 465~ms spin period of \psr.  
We used the {\sc barycen} task and DE405 ephemeris to apply the 
barycentric correction and then extracted filtered for background 
flaring activity events in the 0.2--2~keV band  
from the 22-arcsec radius circle aperture centred at the pulsar position. 
The aperture was selected using the {\sc eregionanalyse} task which
produces the optimum extraction radius basing on signal-to-noise ratio.
This results in 1026 source counts 
or $\approx 98$ per cent of the total number 
of source counts detected in the whole PN energy band.
To search for the pulsations, we performed the $Z^2_n$-test 
\citep{ztest} for the number of harmonics $n$ from 1 to 5 
using the 0.8~mHz window around the spin frequency of 2.150474376(14)~Hz 
expected at the epoch of the \xmm\ observation (MJD~59494) 
according to the \fermi\ timing results (Table~\ref{tab:pars}). 
We detected pulsations at the frequency of 2.150493(2)~Hz 
and found statistically significant contributions from two leading harmonics.
The resulting periodogram is shown in Fig.~\ref{fig:ztest}. 
The maximum $Z^2_2=42.7$, which implies the detection confidence level 
of $\approx 4.7\sigma$\footnote{See Appendix \ref{a:simulations} for details.}.

\begin{figure}
\begin{minipage}[h]{1.\linewidth}
\center{\includegraphics[width=1.0\linewidth,clip]{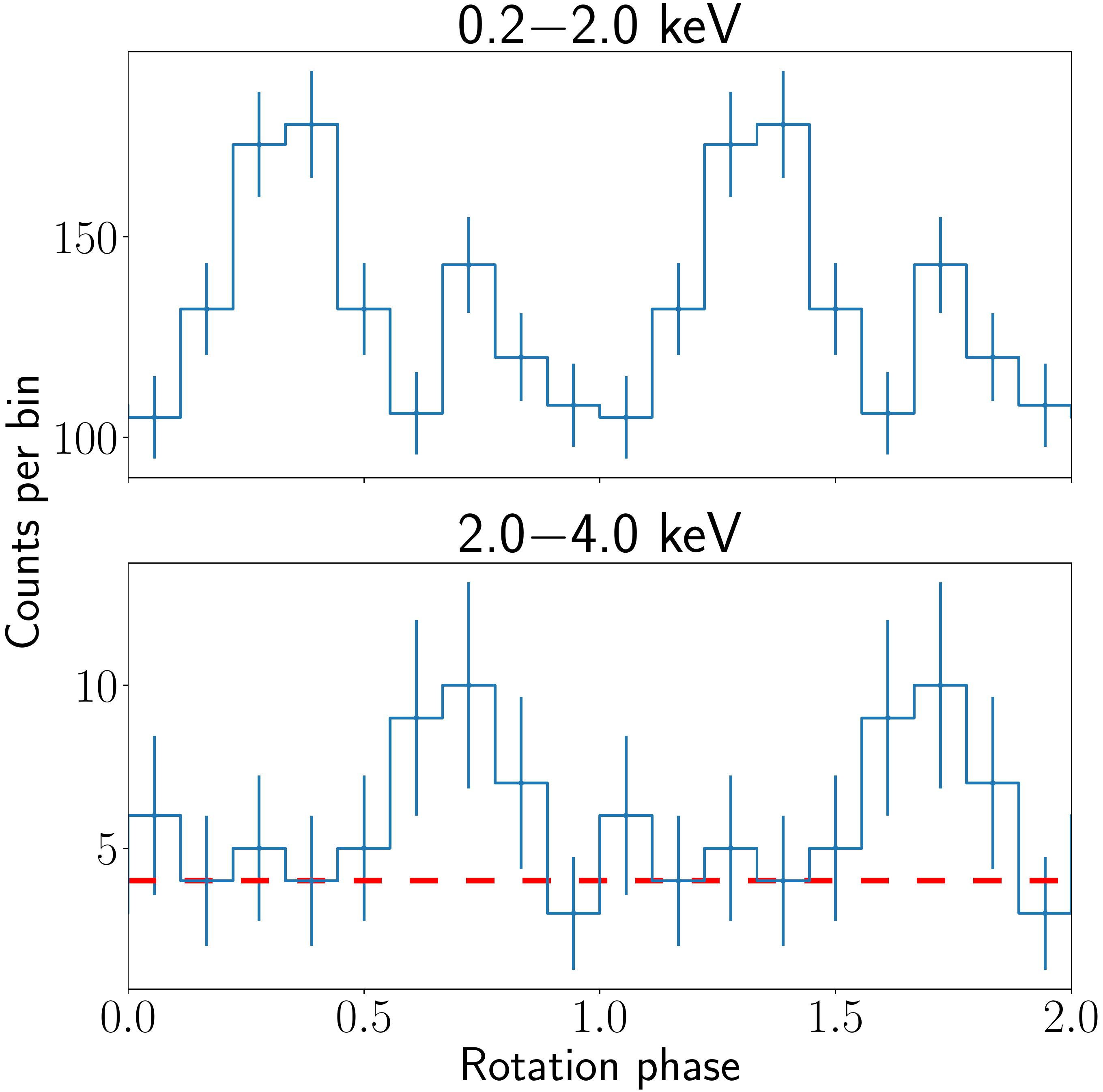}}
\end{minipage}
\caption{Pulse profiles of \psr\ in the 0.2--2~keV and 2--4~keV energy bands. Two phase cycles are plotted for clarity.  
The dashed red line in the bottom panel indicates the background level. 
In the 0.2--2~keV band the background level ($\approx 19$~counts per bin) is significantly below the minimum of 
the pulse profile and the bottom boundary of the plot. 
}
\label{fig:prof}
\end{figure}

We note that the frequency detected in X-rays is somewhat higher 
than the predicted one. This can be due to  
the pulsar timing noise and/or glitches, which might have occurred during the $\sim 8$~yr period 
between the last $\gamma$-ray observations of 2013, included by 
\citet{pletsch2013} into the \fermi\ timing solution, 
and the \xmm\ observations of 2021. For instance, the relative difference $\log(\Delta f/f) \approx -5.1$ is consistent with the distribution of the relative   glitch sizes $\log(\Delta f_\mathrm{g}/f)$  observed for pulsars \citep{Lower2021}.

\begin{table*}
\renewcommand{\arraystretch}{1.3}
\caption{Best-fitting parameters for different spectral models$^\dag$.}
\label{tab:mcmc:bestfit}
\begin{center}
\begin{tabular}{lccccc}
\hline  
Model                              & 2BB + PL               & 2BB + PL$^\ddag$                   & \bbg\                    & \nsl\                    & \nsh\                  \\
\hline
\nh, $10^{21}$ cm$^{-2}$           & $3.1_{-0.2}^{+0.2}$    & $5.2_{-0.8}^{+0.4}$                & $2.3_{-0.4}^{+0.5}$      & $1.62_{-0.06}^{+0.08}$  & $1.66_{-0.07}^{+0.08}$   \\
$D$, kpc                           & $5.5_{-0.5}^{+0.3}$    &                                    & $4.1_{-1.4}^{+1.4}$      & $2.0_{-0.4}^{+0.2}$     & $2.0_{-0.2}^{+0.4}$      \\
$\Gamma$                           & $2.0^\mathrm{fixed}$      & $2.0^\mathrm{fixed}$                  & $2.0^\mathrm{fixed}$        & $2.2_{-0.4}^{+0.6}$     & $2.3_{-0.4}^{+0.5}$      \\
$K$, $10^{-6}$ \phs\               & $2.2_{-0.6}^{+0.6}$    & $1.9_{-0.7}^{+0.6}$                & $2.1_{-0.5}^{+0.6}$      & $1.6_{-0.7}^{+1.3}$     & $2.0_{-0.7}^{+1.5}$      \\
log $L_\mathrm{X}$, \ergs          & $31.27_{-0.15}^{+0.13}$& $29.62_{-0.12}^{+0.20}+\mathrm{log}D_{0.9}^2$& $31.08_{-0.28}^{+0.20}$  & $30.21_{-0.30}^{+0.15}$ & $30.27_{-0.19}^{+0.19}$  \\
\hline
$\alpha$, deg                      &                        &                                    &                          & $70_{-20}^{+20}$        & $60_{-20}^{+20}$         \\
$\zeta$, deg                       &                        &                                    &                          & $60_{-10}^{+20}$        & $80_{-30}^{+10}$         \\
$M$, \msun                         &                        &                                    &                          & $1.9_{-0.2}^{+0.2}$     & $1.8_{-0.2}^{+0.3}$      \\
$R$, km                            &                        &                                    &                          & $13.5_{-1.7}^{+1.2}$    & $13.0_{-1.3}^{+1.5}$     \\
$R^\infty$, km                     &                        &                                    &                          & $16.9_{-1.3}^{+1.5}$    & $17.1_{-1.3}^{+1.3}$     \\
$k_\mathrm{B}T^\infty$, eV            &                        &                                    &                          & $47_{-2}^{+2}$          & $49_{-2}^{+2}$           \\
$R^\infty_\mathrm{cold}$, km          & $19_{-5}^{+1}$         & $14_{-6}^{+5}\ D_{0.9}$            & $19_{-9}^{+1}$           &                         &                          \\
$R^\infty_\mathrm{hot}$, km           & $1.12_{-0.11}^{+1.05}$ & $0.33_{-0.20}^{+0.23}\ D_{0.9}$    & $1.19_{-0.17}^{+0.87}$   &                         &                          \\
$k_\mathrm{B}T^\infty_\mathrm{cold}$, eV & $86_{-4}^{+5}$         & $65_{-3}^{+7}$                     & $84_{-10}^{+6}$          &                         &                          \\
$k_\mathrm{B}T^\infty_\mathrm{hot}$, eV  & $156_{-25}^{+12}$      & $140_{-17}^{+23}$                  & $135_{-13}^{+19}$        &                         &                          \\
log $L^\infty$, \ergs\              & $33.32_{-0.19}^{+0.10}$& $32.79_{-0.39}^{+0.11}+\mathrm{log}D_{0.9}^2$& $33.23_{-0.61}^{+0.12}$  & $32.25_{-0.11}^{+0.12}$ & $32.33_{-0.10}^{+0.10}$  \\
\hline
$E_0$, eV                          &                        &                                    & $370_{-70}^{+30}$        & $340_{-40}^{+40}$       & $350_{-50}^{+30}$        \\
$\sigma$, eV                       &                        &                                    & $24_{-14}^{+18}$         & $25_{-9}^{+13}$         & $24_{-8}^{+14}$          \\
$\tau$, eV                         &                        &                                    & $> 40$                   & $> 680$                 & $> 890$                  \\
EW, eV                             &                        &                                    & $< 430$                  & $150_{-40}^{+120}$      & $180_{-70}^{+70}$        \\
\hline
$W$/\dof                           & $240 / 216$            & $231 / 216$                        & $228 / 213$              & $228 / 211$             & $228 / 211$              \\
\chir/\dof                         & $1.46 / 40$            & $1.16 / 40$                        & $1.14 / 37$              & $1.27 / 34$             & $1.27 / 34$              \\
\hline
\end{tabular}
\end{center}
\begin{tablenotes}
\item $^\dag$ \nh\ is the absorbing column density, $D$ is the distance,
$\Gamma$ is the photon index, $K$ is the PL normalisation, 
$L_\mathrm{X}$ is the non-thermal luminosity in the 2--10~keV band,
$\alpha$ is the angle between the axis of rotation and the magnetic axis,
$\zeta$ is the angle between the rotation axis and the line of sight,
$M$ is the NS mass, $R$ and $R^\infty=R(1+z_\mathrm{g})$ are the intrinsic 
and apparent radii of the NS, 
$T^\infty=T/(1+z_\mathrm{g})$ is the atmosphere redshifted effective temperature, 
$R^\infty_\mathrm{cold}$ and $R^\infty_\mathrm{hot}$ 
are the apparent radii of the equivalent emitting spheres,
$T^\infty_\mathrm{cold}$ and $T^\infty_\mathrm{hot}$ 
are the redshifted effective temperatures of the cold and hot BB components,
$L^\infty=L/(1+z_\mathrm{g})^2$ is the apparent bolometric thermal luminosity, 
$E_0$, $\sigma$, $\tau$ and EW are the absorption line centre, width, 
depth and equivalent width, $z_\mathrm{g}$ is the gravitational redshift. 
All errors are at 1$\sigma$ credible intervals,
while the lower and upper limits are set at 98 per cent. 
The last two rows provide the minimum values of $W$-statistics,
which was used as a log-likelihood in the MCMC procedure,
and \chir\ calculated for the spectra which were grouped 
to ensure at least 25~counts per bin.
\item $^\ddag$ In this case, in contrast to all other models, 
the \nh--$D$ relation was not used in 
the fitting procedure (see text for details). 
$D_{0.9} \equiv D/0.9$~kpc, where 0.9~kpc is the lower limit of 
the distance to \psr\ obtained from its association with the SNR \snr\ 
\citep{zhao2020}.
\end{tablenotes}
\end{table*}

Folded X-ray pulse profile in the 0.2--2~keV energy band is presented 
in the upper panel of Fig.~\ref{fig:prof}. Following the approach suggested
by \citet{becker1999} we calculated the optimal number of phase bins to be 11. 
However, this value exceeds the maximum of 9 bins set by the time resolution of the PN camera 
and the pulsar period, so we used the latter value to construct the folded light curve.
Two peaks separated by about a half of the pulsar rotation phase 
are clearly resolved. The pulsed fraction (PF) of the emission was calculated from 
the photon phases using the bootstrap method outlined in \citet*{swanepoel1996}. 
The result obtained with this technique is not affected by binning effects, 
which is particularly important in our case since the number of counts is low and the 
phase bins are wide. The resulted background-corrected PF in the 0.2--2~keV band is 
$25\pm6$ per cent. In the lower panel of Fig.~\ref{fig:prof} we also show the pulse 
profile in the 2--4~keV energy band. The pulsar possibly demonstrates single-peaked 
pulsations, but the count statistics in this band is too low to make definite conclusions.


\section{Spectral analysis}
\label{sec:spec}

We extracted the pulsar spectra from the 19-arcsec radius aperture 
using {\sc evselect} task. This radius was chosen using the 
{\sc eregionanalyse} task as the optimum value in terms of 
signal-to-noise ratio for the 0.2--10~keV band.
The redistribution matrix and the ancillary
response files were generated by {\sc rmfgen} and {\sc arfgen} tools.
For the background, we chose the 38.5-arcsec radius circular source-free region,
located at the same CCD chip and approximately the same CCD RAWY pixel position as the source 
(see Fig.~\ref{fig:img}), 
in accordance with general recommendations by the EPIC Consortium\footnote{See, e.g., 
footnote \ref{CAL-TN-0018}, pages 28--29}.
As a result, we obtained 221, 246 and 1001 net counts in the 0.2--10~keV band
from the MOS1, MOS2 and PN data, respectively. We fitted the spectra
simultaneously with the X-Ray Spectral Fitting Package ({\sc xspec}) 
v.12.11.1 \citep{arnaud1996} using the {\sc tbabs} model with 
the {\sc wilm} abundances \citep*{wilms2000} to take into account the interstellar medium (ISM)
absorption.

At first, we grouped the spectra to ensure at least 25 counts 
per energy bin and used $\chi^2$-statistics to preliminarily check,  
which of the models, typically used to describe pulsars X-ray emission, can fit the data. 
The single {\sc powerlaw} (PL) model, corresponding to  
the NS magnetosphere emission,     
though statistically acceptable (\chir/\dof~= 1.03/41, \dof~$\equiv$ degrees of freedom) 
resulted in a photon index $\Gamma \approx 7$, which is too high to be consistent 
with typical slopes of non-thermal X-ray spectra of pulsars  
\citep[e.g.][]{kargaltsev2008}. 
Addition of the blackbody (BB) component (model \textsc{bbodyrad} in \textsc{xspec}) 
still gives a too high photon index, 
while the single BB model fits the spectrum worse (\chir/\dof~= 2.02/43).
In contrast, the BB~+ BB model, composed of two BB components with different temperatures and radii of equivalent emitting spheres,
is statistically acceptable with
\chir/\dof~= 1.09/41. 
This model is usually assumed to describe thermal emission from some
colder and hotter areas of the NS surface.
However, there is some flux excess 
above the model at energies $\gtrsim 2$~keV. Although its significance is low, 
this may indicate the presence of the non-thermal emission.
Thus, we added the PL component to the BB~+ BB model, which resulted in 
\chir/\dof~= 1.04/39 and a reasonable photon index $\Gamma \lesssim 3$.

We also constructed the hydrogen atmosphere models {\sc nsmdintb} for 
the NS with a dipole magnetic field to fit the thermal spectral component. 
In these models, which are described in Appendix~\ref{sec:atm}, free parameters are: 
the NS mass $M$ and radius $R$, the distance $D$,
the angle $\alpha$ between the axis of rotation and the magnetic axis,  
the angle $\zeta$ between the rotation axis and the line of sight, 
and the redshifted effective temperature $T^\infty = T/(1+z_\mathrm{g})$, 
where $z_\mathrm{g}$ is the gravitational redshift at the NS surface.
The effective surface temperature $T$ is defined by the total thermal luminosity $L$ 
according to the relation
$L \equiv 4\pi\sigma_\mathrm{SB}R^2 T^4$, where 
$\sigma_\mathrm{SB}$ is the Stefan-Boltzmann constant.
We used two values of the magnetic field at the pole close to the estimates based on the
dipole spin-down formula (see Appendix~\ref{sec:atm}), $B_\mathrm{p}=10^{13}$~G and 
$2\times10^{13}$~G (hereafter NS130 and NS133, respectively). 
We obtained rather poor fits with \chir/\dof~= 1.60/39 for NS130 
and \chir/\dof~= 1.55/39 for NS133. 
However, we found that addition of an absorption Gaussian line
(model \textsc{gabs} in {\sc xspec}) at $\approx 0.35$~keV significantly 
improves the fit resulting in \chir/\dof~= 1.15/36 and 1.17/36. 
Addition of the PL component to describe the flux excess at high energies slightly 
improves the fits 
leading to \chir/\dof~= 1.15/34 and  1.07/34. 
Basing on these preliminary tests,  
we further focused on the models 2BB~+ PL and \nsm\ as the ones providing the best fit statistics.  

\begin{figure}
\begin{minipage}[h]{1.\linewidth}
\center{\includegraphics[width=1.0\linewidth,clip]{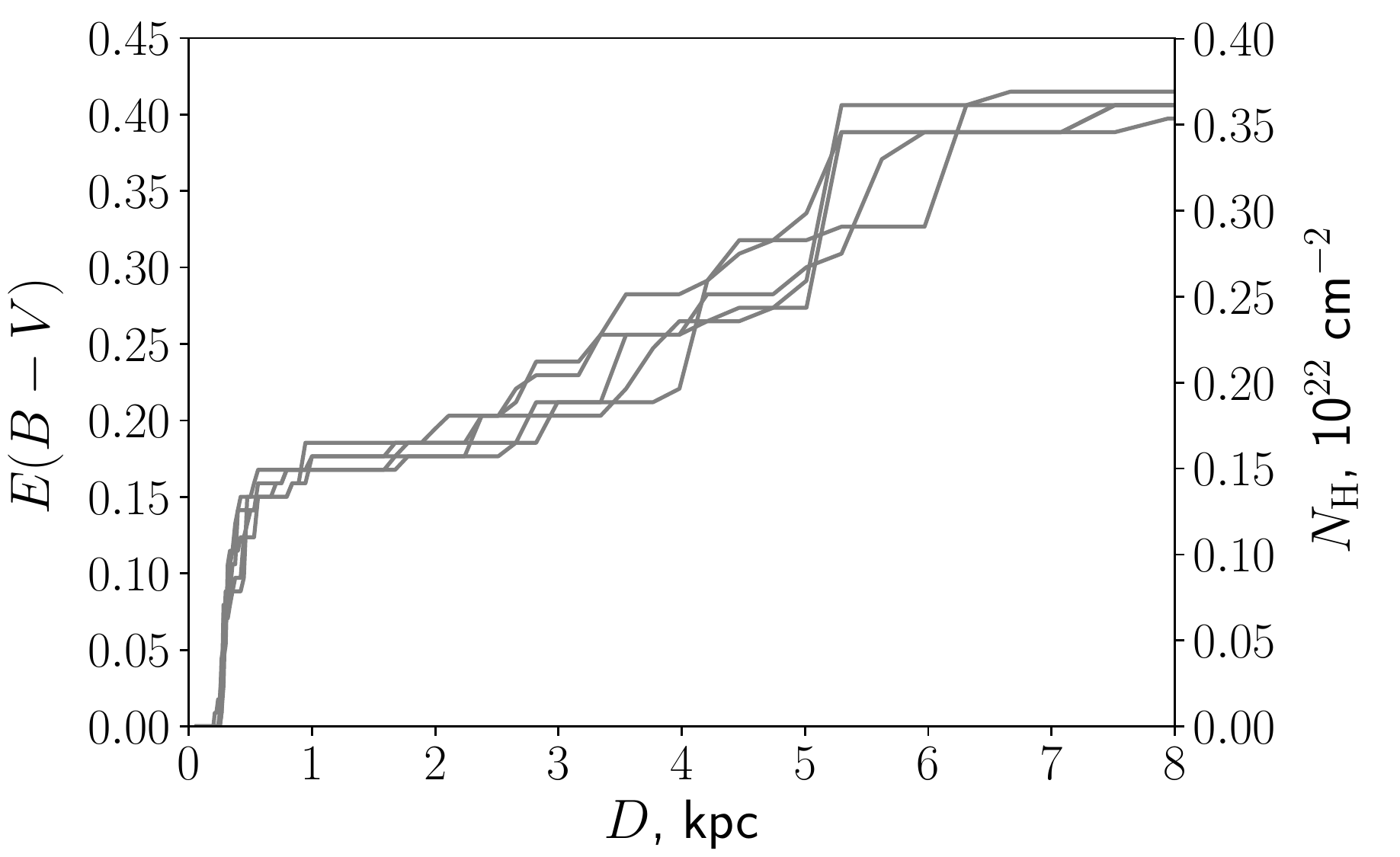}}
\end{minipage}
\caption{Interstellar extinction--distance relation in the \psr\ direction
obtained from the dust map of \citet{green2019}.}
\label{fig:nh-d}
\end{figure}

The number of collected source counts is not large and for more rigorous analysis   
we regrouped the spectra to ensure at least 1 count per energy bin. 
This allows us to obtain the most robust estimates of 
the \psr\ spectral parameters. We applied $W$-statistics \citep*{wachter1979}
appropriate for Poisson data with Poisson background\footnote{See
\url{https://heasarc.gsfc.nasa.gov/xanadu/xspec/manual/XSappendixStatistics.html}.}
and performed the fitting using the Markov chain Monte Carlo (MCMC) 
technique. We utilised the Bayesian parameter estimation procedure using {\sc pyxspec} 
interface and the {\sc python} package {\sc emcee} \citep{emcee2013},  
which implements the affine-invariant MCMC sampler 
developed by \citet{goodman&weare2010}. The best-fitting model parameters, 
defined as the ones corresponding to the maximum values of probability density, were derived from 
the sampled posterior distributions together with their 1$\sigma$ credible 
intervals. 

The Bayesian inference also allows us to include some additional 
information in the fitting procedure.
We used the 3D map\footnote{\url{http://argonaut.skymaps.info/}} 
of the dust distribution in the Galaxy 
based on \textit{Gaia}, Pan-STARRS 1 and 2MASS data 
\citep{green2019} to obtain the extinction--distance
relation in the \psr\ direction. 
In Fig.~\ref{fig:nh-d} we show five representative samples of this relation
drawn from the posterior distribution of the distance--reddening profiles
(see \citealt{green2019} for details). The following procedure was used. 
At each step of the MCMC fitting, we randomly take one of the five samples and 
use the relation \nh\ = $a \times10^{21}E(B-V)$ to 
convert the selective reddening $E(B-V)$ into the equivalent hydrogen column 
density \nh, which is responsible for the ISM absorption in X-rays. 
The conversion factor $a$ is drawn from the normal 
distribution with the mean of 8.9 and the standard deviation of 0.4 according to 
the empirical relation \nh\ = $(2.87\pm0.12)\times10^{21}A_{V}$~cm$^{-2}$ 
\citep{foight2016} and implying the standard reddening law $A_{V} = 3.1 E(B-V)$, 
where $A_{V}$ is the optical extinction. Then we compute the distance $D$ using linear 
interpolation between the closest sample values $\underline{N_\mathrm{H}}$ and 
$\overline{N_\mathrm{H}}$ such that $\underline{N_\mathrm{H}} 
\leq$ \nh\ $\leq \overline{N_\mathrm{H}}$. 

In contrast to {\sc nsmdintb} models, where both the NS radius and the distance are free
parameters, the BB model has a normalisation $N=(R_\mathrm{BB}^{\infty}/D)^2$ as a free parameter,
where $R_\mathrm{BB}^{\infty}$ is the apparent radius of the equivalent emitting sphere.
Thus, in the case of the 2BB~+ PL model at each step we independently 
sampled the radii $R^{\infty}_\mathrm{cold}$ and $R^{\infty}_\mathrm{hot}$ for colder and hotter emitting areas 
and then calculated normalisations 
using these values together with the computed distance. 
We constrained both radii to be $\leq 20$~km as a reasonable
value of an NS radius measured by a distant observer.

For the \nsm\ models, we used a prior $\alpha+\zeta~\geq 90\degs$,  
which follows from the shape of the folded light curve and the PF
value.
The PL photon indices $\Gamma$ were constrained in the range of 0.5--3 
typical for pulsars \citep[e.g.][]{kargaltsev2008}. However,
in the case of the 2BB~+ PL model, $\Gamma$ tends to the upper bound, 
while the temperature of the hotter BB component in general takes slightly
lower values than that in the case of the pure BB~+ BB model. This behaviour is 
typical for situations when the PL component competes with the hotter 
BB component at lower energies, while the number of counts above $\sim 2$~keV 
is very low and the PL slope there is poorly constrained. This is 
exactly the case of J0554. Thus, we applied the fixed photon index 
$\Gamma=2$ for the 2BB~+ PL model. 

\begin{figure}
\begin{minipage}[h]{1.\linewidth}
\center{\includegraphics[width=1.0\linewidth,clip]{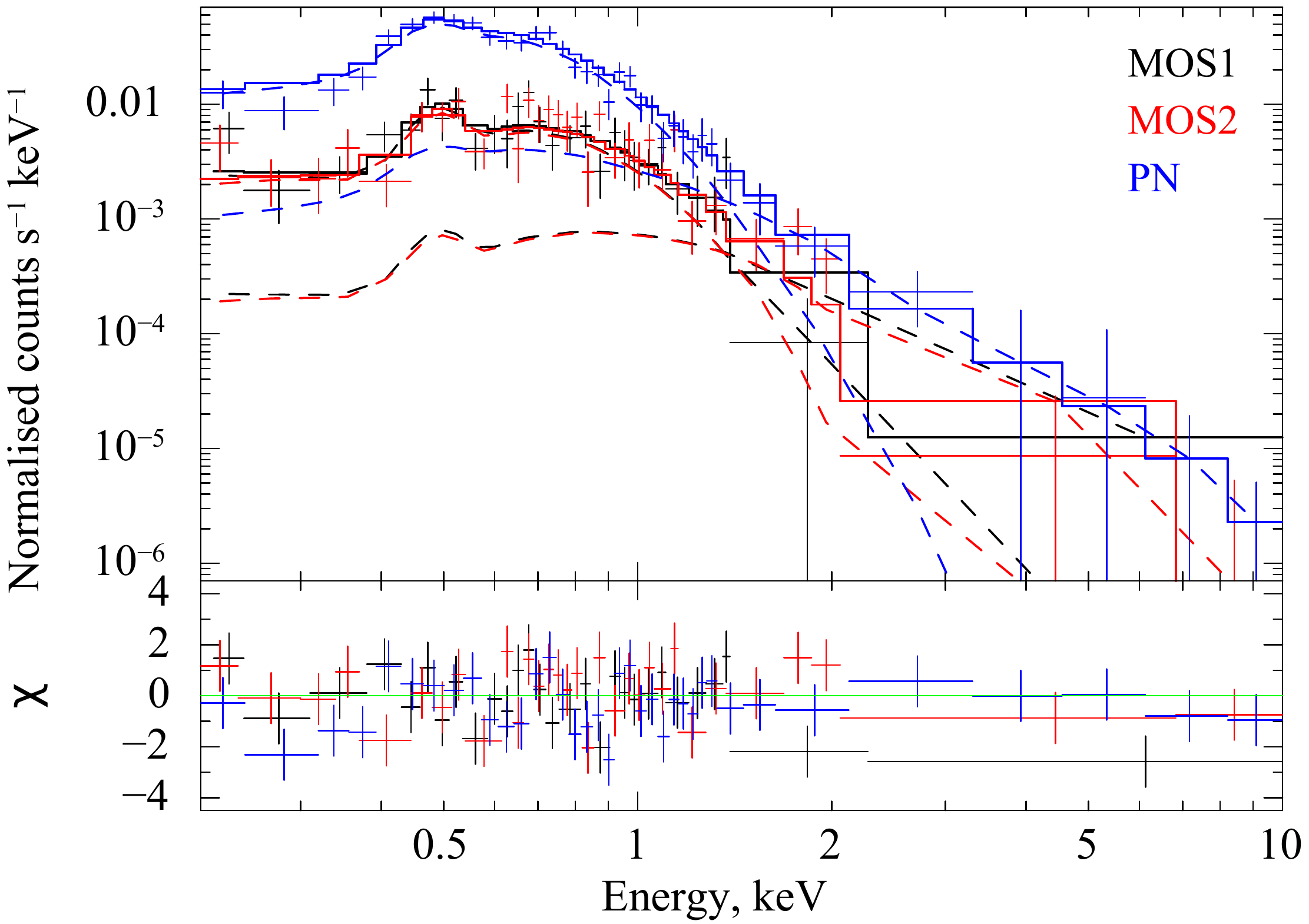}}
\end{minipage}
\caption{\psr\ spectrum, best-fitting model \nsh\ and residuals. 
The dashed lines show the model components. Data from different detectors are
shown by different colours. For illustrative purposes, spectra were regrouped
using {\sc xspec} command {\sc rebin}
to reach at least 2$\sigma$ detection significance in each new bin 
(but no more than 12 adjacent bins were allowed to be combined).}
\label{fig:spec}
\end{figure}

\begin{figure*}
\includegraphics[width=1.99\columnwidth,clip]{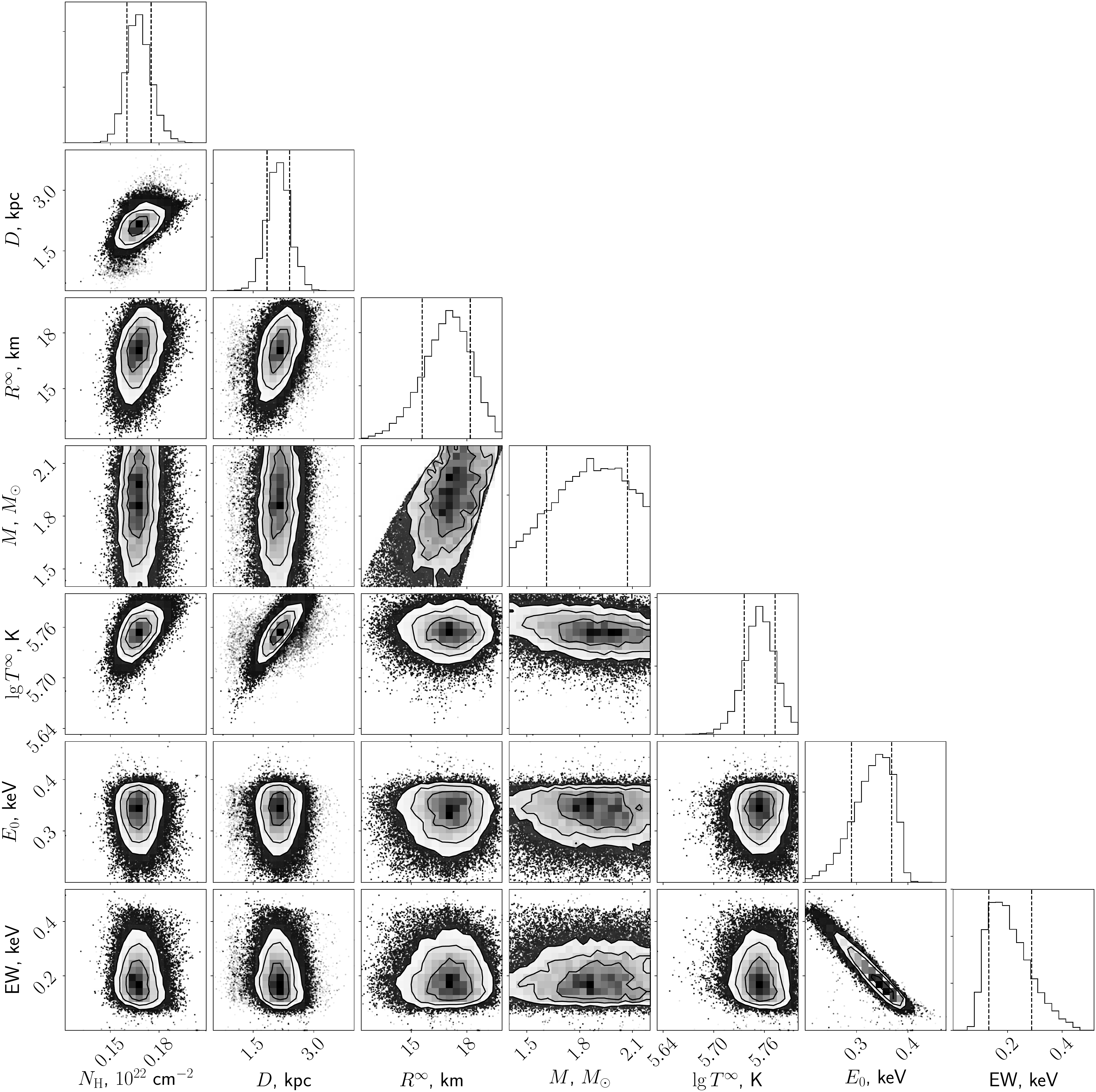} 
\caption{1D and 2D marginal posterior distribution functions for some parameters of the \nsh\ model. Contours in 2D distributions correspond to 39, 68 and 86 per cent confidence levels. In 1D distributions, vertical dashed lines indicate 1$\sigma$ credible intervals.
\label{fig:triangle}}
\end{figure*}

The best-fitting parameters are presented in Table~\ref{tab:mcmc:bestfit}. 
The pulsar spectrum with the best-fitting model \nsh\ 
is shown in Fig.~\ref{fig:spec}. In Fig.~\ref{fig:triangle} we present a corner plot of 
posterior distribution functions for some parameters of this model. To check the fit quality,
we calculated values of the $\chi^2$-statistics using 
the spectra grouped to ensure at least 25 counts per bin.  
They are provided in the last row of Table~\ref{tab:mcmc:bestfit}. 
These $\chi^2$ values differ from the preliminary ones partially due to 
inclusion of parameter constraints in the fitting procedure.

The Gaussian line model \textsc{gabs} describing the spectral feature is represented as
\begin{equation}
    \mathfrak{G}(E) =\mathrm{exp}\left(-\frac{\tau}{\sqrt{2\pi} \sigma} e^{-\frac{(E-E_0)^2}{2\sigma^2}} \right),   
\end{equation}
where $E_0$, $\sigma$ and $\tau$ are the line centre, width and depth.
$\tau$ is poorly defined from the spectral fits and we provide
only its lower limits. We also calculated the line equivalent 
width (EW) defined as:
\begin{equation}
    \mathrm{EW} = \int (1 - \mathfrak{G}(E))\mathrm{d}E,    
\end{equation}
which is better constrained by the fits.

From the 2nd column of Table~\ref{tab:mcmc:bestfit}, one can see that the 2BB~+ PL model
fails to fit the data when the \nh--$D$ relation is used. 
We have found that it describes the spectrum poorly at energies $\lesssim 0.4$~keV, 
where the model flux becomes systematically higher than the observed one. 
Exclusion of the \nh--$D$ prior removes the discrepancy in the soft band and 
results in the statistically acceptable fit (the 3rd column of Table~\ref{tab:mcmc:bestfit}). 
However, for reasonable BB normalisations, 
which imply the apparent radii of $\lesssim 20$~km and the distances of $\gtrsim 0.9$~kpc, 
the obtained column densities are about 1.5 times higher than the maximum 
value provided by the dust map of \citet[][see Fig.~\ref{fig:nh-d}]{green2019}.
Recalling that the addition of the Gaussian absorption line significantly improves
the fit in the case of the atmosphere models, we tried to add the line to the 2BB~+ PL 
model as well. This leads to the good fit without \nh\ anomalies even when 
the \nh--$D$ 
relation is implemented (see the 4th column in Table~\ref{tab:mcmc:bestfit}).

\section{Discussion}
\label{sec:discussion} 

\subsection{\psr\ spectral and timing properties}
\label{subsec:spec&tmg}

The time-integrated X-ray spectrum of \psr\ can be well described  
by the composite emission models containing the thermal and non-thermal spectral components.
For the former, we tried the BB model, which imitates the emission 
from the solid or liquid  state NS surface, and the hydrogen atmosphere models with 
dipole magnetic fields. 

In the case of the 2BB~+ PL model without the \nh--$D$ prior, we assume that
the cold BB component originates from the bulk of the stellar surface 
while the hot component describes emission from a hot spot (e.g. polar caps). 
Then the distance to \psr\ should be $\lesssim 2$~kpc, otherwise 
the radius of the cold component will be inadequately large (see
the 3rd column of Table~\ref{tab:mcmc:bestfit}). However, the model requires 
a much higher column density than expected for such distance according to the
dust map by \citet{green2019}. This means that either the model, 
being formally statistically acceptable, results in a physically 
inappropriate \nh\ and hence wrong NS parameters, or the considered 
\nh--$D$ relation is incorrect in the \psr\ direction. 
Since the pulsar is projected onto the SNR \snr, the derived absorption 
excess could be provided by some filament of the remnant even if it is 
not visible in X-rays. Unfortunately, there are no X-ray  
extragalactic objects in the pulsar field which
are bright enough to make an independent estimate of the maximum absorption 
in the pulsar direction. However, we checked some other extinction maps 
by \citet*{drimmel2003}, \citet{chen2014}, \citet{chen2019}, 
\citet{sale2014} and \citet{lallement2014,lallement2018,capitanio2017}.
All of them provide $N_\mathrm{H} \lesssim 3.5\times 10^{21}$~cm$^{-2}$ 
at 2~kpc which is significantly smaller than the best-fitting value. 
Moreover, basing on the [O~{\sc iii}]/H$\alpha$ emission ratio measured
for the SNR \snr\ associated with \psr, \citet{how2018} argued 
that $E(B-V)$ should not be much
greater than 0.3 (i.e. \nh~$\lesssim2.7\times10^{21}$~cm$^{-2}$). 
This makes the 2BB~+ PL model  hardly acceptable 
without additional constraints and model components. 

Implementation of the \nh--$D$ prior along with the addition 
of the low energy absorption feature 
solves the 2BB~+ PL problem, 
resulting in about twice lower column density and acceptable fit statistics.
If we assume that the cold BB component describes the emission 
from the bulk of the stellar surface while the hot component corresponds 
to a hot spot, then the size of the latter is larger than the `standard' polar cap 
of a radius of a few hundred meters, expected according to the model of \citet{sturrock1971}. 
The obtained effective temperatures are in agreement with results for other 
NSs of similar age \citep{potekhin2020}.

The \nsm\ models are also plausible. Parameters obtained 
for two different magnetic fields are very similar
(see the last two columns of Table~\ref{tab:mcmc:bestfit}).\footnote{We also tried analogous models with other field strengths and found that 
a decrease of $B_\mathrm{p}$ to $\sim 2\times10^{12}$~G worsens the fitting statistics, while an increase to $5\times10^{13}$~G does not noticeably change the results.} 
These models indicate that  
\psr\ should be a rather heavy NS, 
with the mass of 1.6--2.1~\msun. Its redshifted effective temperature 
$T^\infty = 48 \pm 3$~eV ($0.56 \pm 0.3$~MK), which is about twice 
lower than the temperature of the cold BB component from the bulk surface of the NS 
in the 2BB~+ PL model. This is a typical situation when the thermal NS component 
is equally well described by the BB and atmosphere models, since the latter 
spectra are harder \citep{potekhin2014}. 

It is important, that all three statistically acceptable models 
implementing the \nh--$D$ prior require the absorption feature 
at $\approx 0.35$~keV  regardless the local continuum shape. 
This supports the presence of the feature in the data.   
In the 2BB~+ PL model without the \nh--$D$ relation, 
which we consider as hardly acceptable, the absence of the absorption line is compensated by 
the implausibly  high \nh\ value. 
We note that there are only a few rotation powered pulsars for 
which absorption lines have been reported: 
PSR J1740$+$1000 \citep{kargaltsev2012}, 
PSR J0659$+$1414 \citep{arumugasamy2018,zharikov2021,schwope2021}, 
PSR J0726$-$2612 \citep{rigoselli2019},
PSR J1819$-$1458 \citep{mclaughlin2007,miller2013} and
Calvera \citep{shevchuk2009,shibanov2016,mereghetti2021}. 
For none of these objects the nature of lines has been unambiguously established. 

The nature of the \psr\ feature is also unclear. One of possible explanations is
a cyclotron absorption line. Such line position measured by 
a distant observer is given by:
\begin{equation}
    E_\mathrm{cyc}^\infty = 11.577\ (1+z_\mathrm{g})^{-1} Z\ \frac{m_\mathrm{e}}{m} \frac{B}{10^{12}\ \mathrm{G}}\ \mathrm{keV},    
\end{equation}
where $m_\mathrm{e}$ is the electron mass,
$Z$ and $m$ are the charge number and mass
of the particle that is responsible for the cyclotron absorption. Hence, the surface
magnetic field is $\approx 4\times10^{10}$~G if the line is produced 
by electrons, and $\approx 7\times10^{13}$~G if it is produced by protons or $\approx (3-4)\times10^{13}$~G 
if it is produced by heavier ions.
The first value is $\sim 200$ times lower and the second 
is about an order of magnitude higher than the estimated 
spin-down magnetic field (see Table~\ref{tab:pars} and Appendix~\ref{sec:atm}).
Thus, the electron cyclotron line can be created at a few stellar radii 
above the NS surface, e.g. in a radiation belt, where the magnetic field 
is much weaker \citep[see e.g.][]{luo&melrose2007}. 
Otherwise, if it is an ion cyclotron line, then
the magnetic field at the surface is 
considerably stronger than the $B$ values estimated in Appendix~\ref{sec:atm} and used in our atmosphere modelling.
Such a cyclotron line might indicate the presence of strong multipole
field components \citep[cf.][]{bilous2019,lockhart2019}
or magnetic loops  
\citep*{Tiengo_13,MereghettiPM15,Rodriguez_Castillo_16}, which are not considered in our models. 
We also note that 
magnetized atmosphere models predict too low equivalent width of
the proton cyclotron line in comparison with the observed one
(cf.{} the discussion of a similar case by \citealt{Hambaryan_17}).
In particular, our dipolar models
show that the proton cyclotron line is damped by the smearing due to
magnetic field variations over the surface (see Figs.~\ref{fig:spec_theor}). 

Alternatively, the feature
may be formed by atomic transitions in a non-hydrogen atmosphere
(e.g., \citealt{mori2007}), the ISM or a cloud near the outer part of the
magnetosphere \citep[cf.][]{Hambaryan_09,Pires_19}.

Finally, we cannot exclude the possibility that the absorption feature is
an instrumental artefact. The addition of the line 
improves significantly the fit of the PN spectrum but does not 
influence the statistics for the MOS data. However, this is not surprising
since the former spectrum contains much more counts. 
To check whether it is  the artefact, it would be useful to examine whether 
the spectra of other sources in the PN FoV show similar features. 
Unfortunately, all other sources are not bright enough to perform such analysis.
Phase-resolved spectral analysis of \psr\ could also help 
to clarify the nature of the feature in its emission.   
However, this is impossible because of the low count statistics. 

The non-thermal X-ray luminosity of \psr\ in the 2--10~keV band
$L_\mathrm{X} \approx (1.6-1.9)\times10^{30}$~\ergs\ for the \nsm\ models 
and $1.2\times10^{31}$~\ergs\ for the \bbg\ model. The corresponding 
X-ray efficiencies $\eta_\mathrm{X}=L_\mathrm{X}/\dot{E}\approx10^{-4.8}$--$10^{-4.9}$
and $10^{-3.7}$, respectively\footnote{For the \nsm\ models, we 
recalculated the spin-down luminosity using the best-fitting 
parameters and the formula for the moment of inertia by \citet{RavenhallPethick94}.
The resulting $\dot{E}=(1.2-1.4)\times10^{35}$~\ergs.
For the \bbg\ model, we used $\dot{E}$ from Table~\ref{tab:pars}.}. 
For a 52~kyr old pulsar these values are compatible with  
the dependencies  $L_{X}(t_\mathrm{c})$ and $\eta_\mathrm{X}(t_\mathrm{c})$ based on observations 
of other X-ray emitting pulsars  \citep[see e.g.][]{zharikov&mignani2013}.

Using the \nh--$D$ relation, we found the distance to \psr\ to
be 1.6--2.4~kpc in the case of the atmosphere models. This is 
compatible with the `pseudo-distance' of 1.9~kpc based on the $\gamma$-ray data. 
If this estimate is correct, the SNR \snr\ is located somewhat 
closer than relations between the radio surface brightness and the diameter 
of the SNR predict. On the other hand, the \bbg\ model resulted
in the larger distance of 2.7--5.5~kpc. 
This agrees with the upper limit on 
the distance to \snr\ of about 5~kpc provided by \citet{how2018}. 

\begin{figure}
\includegraphics[width=\columnwidth]{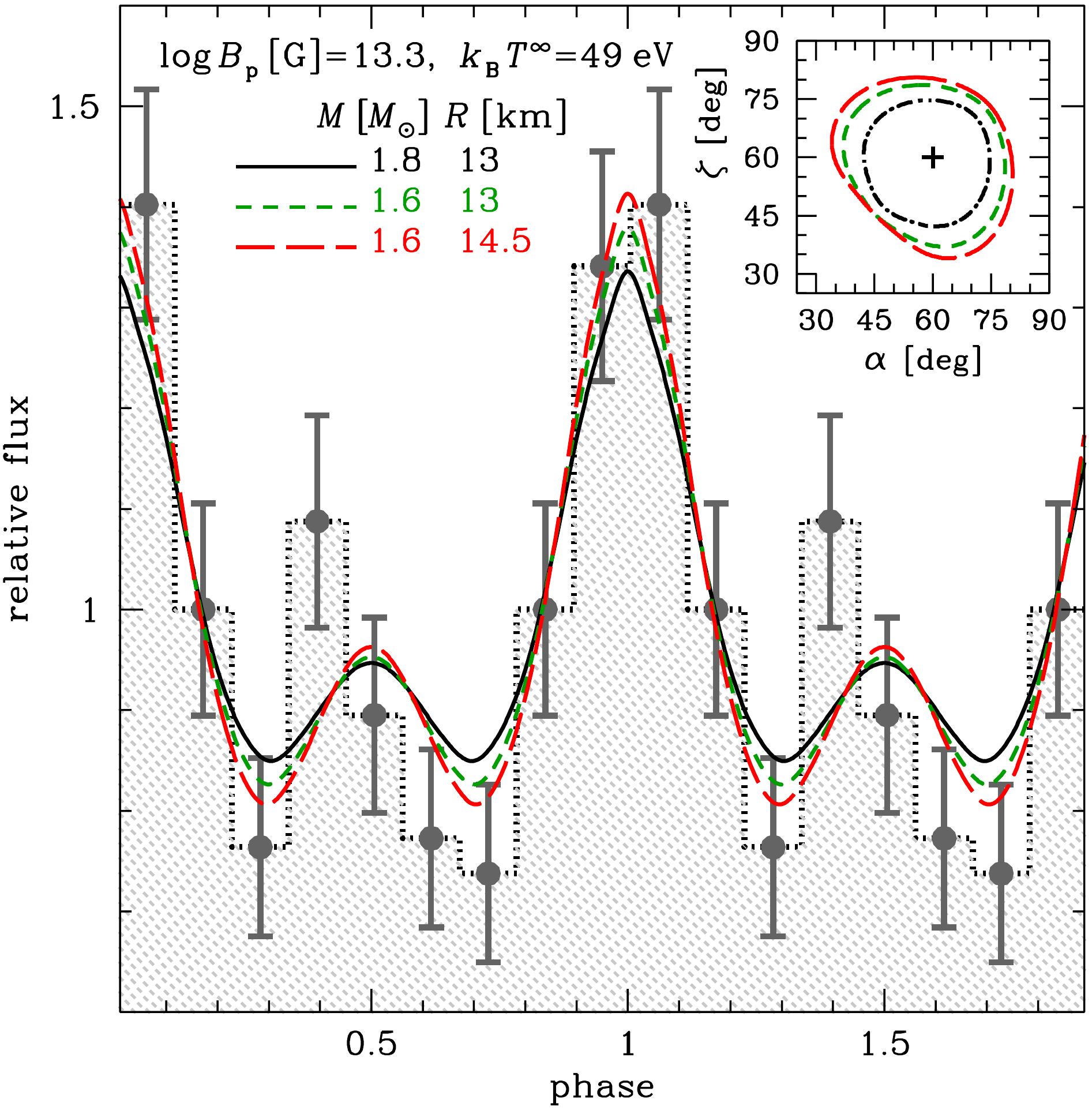} 
\caption{\psr\ pulse profile in the 0.2--2~keV band (grey dotted histogram with error bars) 
and theoretical light curves computed for the model NS133 with
$\alpha=\zeta=60^\circ$, $k_\mathrm{B}T=49$~eV, and for three
different combinations of the mass $M$ and radius $R$ of the NS,
drawn using different line styles as specified in the legend, which 
cover the range of pulsed fractions from 
$\approx 15$ (black solid line) to $\approx 19.5$ per cent (red dashed line).
The inset shows the contours at the 90 per cent confidence level on
the ($\alpha$,$\zeta$) plane for the same three NS models.
\label{fig:lc-fit}}
\end{figure}

We detected, for the first time, X-ray pulsations with the \psr\ spin
period. The pulse profile shows two peaks per period. The pulsed 
fraction in the 0.2--2~keV band is $25\pm6$ per cent. 
This is a typical value for the thermal emission originated from the bulk 
of an NS surface \citep[e.g.][]{pavlov&zavlin2000}. 
Using the pulse profile, we can set additional constraints on the angles 
$\alpha$ and $\zeta$ in the case of the atmosphere models.
The \psr\ observed and theoretical light curves 
are shown in Fig.~\ref{fig:lc-fit}. The maximum PF provided 
by the models {\sc nsmdintb} in the 0.2--2~keV band
is $\approx 20$ 
per cent. This is somewhat lower than the measured value
but is compatible with it within uncertainties. The corresponding 
$\alpha$ and $\zeta$ both lie in the range of 50\degs--70\degs\ 
which is compatible with the results of the spectral analysis
(Table~\ref{tab:mcmc:bestfit}). Despite the low number of counts, 
there is a marginal peak in the 2--4~keV pulse profile which coincides
with the smaller peak in the 0.2--2~keV band (see Fig.~\ref{fig:prof}).
This may indicate the pulsations of the non-thermal component.
As can be seen from Fig.~\ref{fig:spec}, its flux  
in the 0.2--2~keV band is significantly lower than that of 
the thermal component. Nevertheless, it may contribute 
to the pulsed flux and thus somewhat increase the model predicted PF. 
It may also be responsible for some asymmetry of the pulse profile. 
The inset in Fig.~\ref{fig:lc-fit} shows the 90 per cent confidence contours of the 2D distribution of the angles $\alpha$ and $\zeta$ at fixed values
of $M$, $R$, $B$ and $T^\infty$. These contours are obtained assuming that
the flux values, derived from the number of counts in the 9 phase bins, are normally distributed and uncorrelated.

In contrast to the atmosphere models,
the light curves for the 2BB model are almost flat.
This is because the BB components assume uniform temperature distribution 
over the emitting areas and isotropic radiation 
(unlike the peaked radiation from the magnetized atmospheres), but 
mainly because of the smallness of the ratio of the fluxes from the
hot and cold BB components. 
Taking the most probable values for the model \bbg{} from Table~\ref{tab:mcmc:bestfit},
we obtained the strict upper limit $\mbox{PF} < 3.2$ per cent, which is
incompatible with the observed one.


\subsection{\psr\ and NS cooling theories}
\label{subsec:cooling}

\begin{figure*}
\includegraphics[width=\columnwidth]{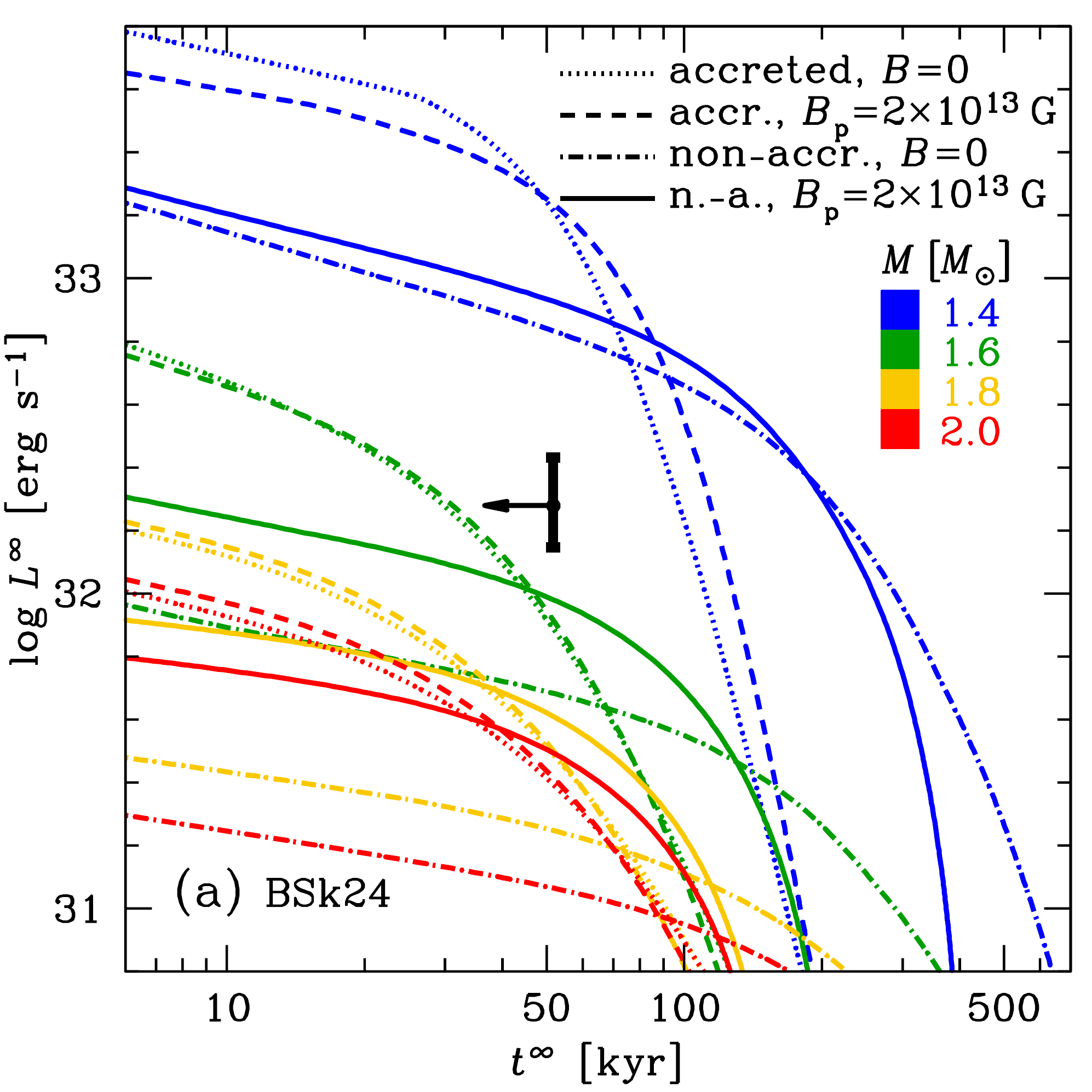}
\includegraphics[width=\columnwidth]{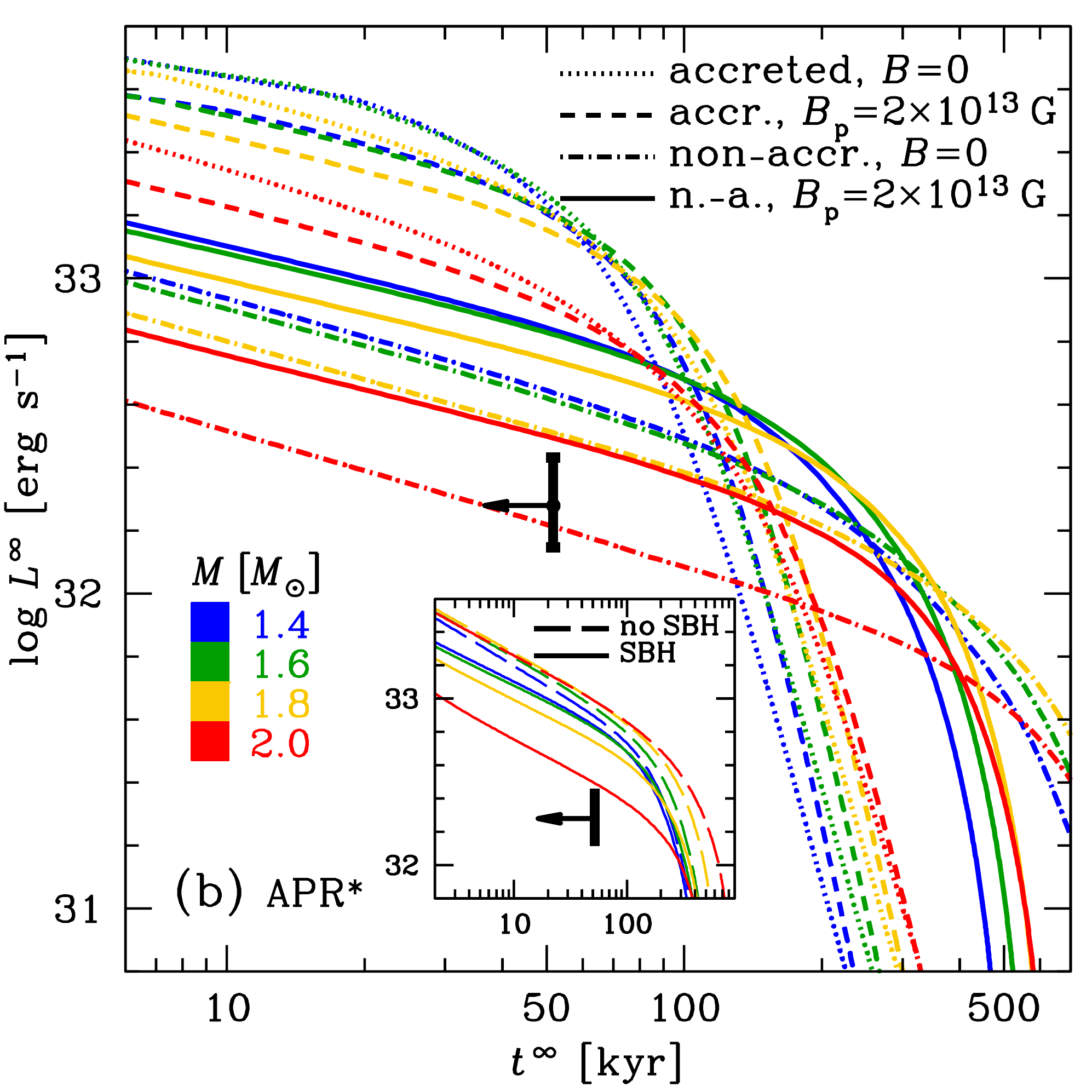}
\caption{Theoretical cooling curves, calculated according to the EoS
models BSk24 (panel a) and APR$^*$ (panel b), compared with
observational estimates of the thermal luminosity and age of J0554 (the error bar). 
The
cooling curves for NS models with masses $M=1.4$, 1.6, 1.8 and
$2.0~\msun$ (coded by color), endowed with a dipole magnetic field having
the strength $B_\mathrm{p}=2\times10^{13}$~G at the pole, are shown by solid
and dashed lines, respectively, for non-accreted and accreted heat blanketing
envelopes (see text for details). For comparison,
cooling curves for analogous non-magnetic NS models are shown,
respectively, by dot-dashed and dotted lines.
The inset to panel b illustrates the role of the enhancement of the modified Urca processes described by \citet*[][ SBH]{Shternin_18}. Here, the solid and long-dashed cooling curves are computed,
respectively, with and without taking the SBH effect into account.
}
\label{fig:cool}
\end{figure*}

The comparison of the measured thermal luminosity with predictions  of
NS cooling theories can provide constraints on the EoS and other
properties of the NS matter and tests for theoretical models of such
matter. In Fig.~\ref{fig:cool} we demonstrate examples of such
comparison. Here, the cooling curves, which show the time-dependence of
the bolometric thermal luminosity of the star in the reference frame of
a distant observer, $L^\infty(t^\infty)$, have been produced using the
computer code presented by \citet{PotekhinChabrier18} with essentially
the same microphysics input. The most uncertain microphysics ingredients
that can significantly affect the cooling are the EoS and composition of
the neutron star core, composition of the heat-blanketing outer
envelopes, and density-dependences of the critical temperatures of the
baryon superfluidity in the core (see, e.g., \citealt*{PotekhinPP15},
for review). Here we show the cooling curves for two EoSs and for two
models of the heat-blanketing envelopes composition. For the EoS and
composition of the core we employ two widely used models: the model
BSk24 \citep{Pearson_18} and the model A18+$\delta$v+UIX$^*$
\citep*{AkmalPR98}, which is hereafter named APR$^*$ and parametrized
according to \citet{PotekhinChabrier18}. Both EoSs describe the $npe\mu$
matter, which consists of the leptons and nucleons without allowance for
other baryons or free quarks. For the composition of the non-accretred
crust and envelopes, we use the BSk24 model of the ground-state matter.
The alternative model is the accreted envelope with helium filling the
heat-blanketing layers up to the density $\rho=10^9$~\gcc{}
\citep*[cf.][]{BeznogovPY21} and with the ground-state composition at
higher densities. We use the model of a magnetized envelope with the
surface distribution of magnetic field and local effective temperature
consistent with the atmosphere model described in
Appendix~\ref{sec:atm}, with the polar magnetic strength
$B_\mathrm{p}=2\times10^{13}$~G. We assume that the field is a relativistic
dipole not only in the atmosphere but also in the stellar core.
Accordingly, we take into account the effects of Landau quantization on
the EoS and thermal conductivities in the outer crust and envelopes and
the heat loss due to the synchrotron and fluxoid-scattering mechanisms
of neutrino emission, as well as magnetically induced modifications of
other neutrino emission mechanisms (see \citealt{PotekhinPP15} and
references therein). For comparison, for each EoS and each envelope
composition, we also plot a cooling curve for the same NS without
magnetic field.

The baryon superfluidity in the $npe\mu$ matter of an NS can be of three
types, characterized by the neutron singlet or triplet and proton
singlet pairing. The theory of the neutron singlet superfluidity, which
occurs in the inner crust of an NS, is sufficiently robust
\citep[see][]{Ding_16}; specifically, here we use the MSH model
\citep{MargueronSH08} in the parametrized form by \citet{Ho_ea15}. In
contrast, there are several substantially different theoretical models
for the other two types of superfluidity that operate in the NS core and
considerably affect the cooling (see, e.g.,
\citealt{Page_14,SedrakianClark19}, for review). For illustration, we
use the parametrizations from \citet{Ho_ea15} for the BS
\citep{BaldoSchulze07} and TTav \citep{TakatsukaTamagaki04} models of
the proton singlet and neutron triplet superfluidity types,
respectively. 

The error bar in Fig.~\ref{fig:cool} embraces the 1$\sigma$ limits to 
the measured thermal luminosity $L^\infty$ for the atmosphere models (the last two columns of
Table~\ref{tab:mcmc:bestfit}). It is placed at $t^\infty=t_\mathrm{c}$, and the
leftward arrow indicates that the true age of the pulsar is likely
somewhat smaller. This expectation is based on the fact that usually
(although not always) true ages of pulsars are smaller than their
characteristic ages (see, e.g., \citealt{potekhin2020}).

Figure~\ref{fig:cool}a shows the cooling curves for the BSk24 model.
These cooling curves pass above the observational error bar for the NS
model with $M=1.4\msun$, but below it for the models with
$M\geq1.6\msun$. The fast cooling of the massive NSs is mainly due to
the powerful direct Urca processes of neutrino emission, which operate
at sufficiently high densities in the cores of these stars and overpower
the more common modified Urca processes. This cooling enhancement occurs
if $M$ exceeds a certain threshold value \citep[e.g.,][]{Haensel95},
which is slightly below $1.6\msun$ for the BSk24 model
\citep{Pearson_18}. The observations can be made compatible with the
enhanced cooling models, if we assume that the true age of J0554 is
smaller than \tc. Among the cooling curves shown in
Fig.~\ref{fig:cool}a, the smallest discrepancy between the true and
characteristic ages is required for the NS with $M=1.6\msun$ and
accreted envelope. In this case $M$ is only slightly larger than
$M_\mathrm{DU}$, so that the direct Urca processes operate only in a
small central part of the NS core.

Figure~\ref{fig:cool}b shows the cooling curves for the APR$^*$ model.
For this EoS, the direct Urca processes cannot occur for $M\lesssim 2\msun$.
However, the non-enhanced (so-called minimal) cooling can be compatible
with the observations of J0554 in this case, if the blanketing envelope
is non-accreted. It is worthwhile to note that the latter compatibility
is achieved due to the enhancement of the modified Urca processes,
recently discovered by \citet*{Shternin_18}. This enhancement becomes
very strong near the threshold density for opening the direct Urca
process. For the APR$^*$ model, the threshold mass only slightly exceeds
$2\msun$, therefore the effect of \citet{Shternin_18} significantly
enhances the total neutrino luminosity of the NS with $M=2\msun$ and
thus decreases its temperature. In the absence of such enhancement,
the luminosity would be higher for NS models with higher masses in the minimal cooling scenario, 
as illustrated by the long-dashed curves in the inset to Fig.~\ref{fig:cool}b, 
because more massive stars have larger heat capacities. 

The presented analysis is self-consistent in the sense that cooling scenarios for 
BSk24 and APR$^*$ models are considered taking into account the NS parameters (mass, 
radius, magnetic field) obtained from the spectral fit. In order to give preference to 
one of the EoSs, better constraints on the NS mass and true age are required.


\section{Summary}
\label{sec:sum} 

Using 45-ks \xmm\ observations we detected a soft point-like X-ray source
within 1$\sigma$ area of the \fermi\ position of the \gr\ pulsar \jpsr,
confirming the earlier \sw\ detection at much higher significance. 
We firmly established the pulsar nature of the source by detecting 
coherent pulsations with the \psr\ frequency in the 0.2--2~keV band. The pulse 
profile demonstrates two peaks separated by about a half of the rotation phase. The 
background-corrected pulsed fraction is $25\pm6$ per cent. Marginal single-peaked 
pulsations are seen in the hard band above 2~keV, but low number of counts 
precludes definite conclusions.

The spectral analysis shows that the thermal emission from the surface of the NS 
dominates at energies below $\approx 2$~keV, while a weak non-thermal magnetospheric 
component may be present in the hard band. To describe the former, we constructed 
a set of the NS hydrogen atmosphere models with dipole magnetic field. In order to 
fit the data they require addition of the absorption line at $\approx 0.35$~keV. 
The nature of the feature is unclear. Among the possibilities are cyclotron 
absorption, atomic transitions in the ISM or NS atmosphere, or an instrumental 
artefact. Implementing the absorption column density--distance relation, we estimated 
the distance to the pulsar to be $\approx 2$~kpc. We note that the combination of 
two blackbody components, corresponding to the cold NS surface with a hot spot on it, 
and the absorption line 
also provides statistically acceptable fit, resulting in about twice as large distance. 
However, this model is less physically realistic and it fails to reproduce 
the observed pulse profile with the pulsed fraction $\gtrsim 20$ per cent. 
Therefore, we consider the atmosphere models as more appropriate ones.

The best fitting parameters obtained for the atmosphere models suggest that \psr\ 
is a rather heavy NS with the mass in the range of 1.6--2.1~\msun\ and the radius 
of about 13~km. The redshifted effective temperature of $\approx 50$~eV corresponds 
to the bolometric luminosity of $\approx 2\times10^{32}$~\ergs\ as seen by a distant observer. 
Utilising this value, together with the pulsar characteristic age of 50~kyr, 
we investigated cooling scenarios for \psr\ in the frame of two popular EoSs. 
For the BSk24 model, the observed bolometric 
luminosity is consistent with the predictions if the mass of the NS is close to the 
lower limit of the range obtained from the spectral fit, and if the true age of \psr\ 
is substantially smaller than the characteristic one. This model also favours the presence 
of the accreted heat-blanketing envelope. On the other hand, the APR$^*$ model requires 
non-accreted envelope and the mass close to 2.0\msun, which leads to effective cooling of 
the NS through modified Urca processes and compatibility of the model cooling curve with 
observations if the true age of \psr\ is close to the characteristic one. 

Deeper X-ray observations are necessary to better constrain the \psr\ parameters.
Phase-resolved spectral analysis would allow one to unveil the nature 
of the absorption feature and to better constrain the pulsar geometry. 
Measurement of the pulsar proper motion could confirm the association
of \psr\ and SNR \snr\ and provide an independent 
estimate of their age. 

\section*{Acknowledgements}
We are grateful to V. Suleimanov for providing calculations of specific 
spectral fluxes from hydrogen atmospheres of NSs with strong magnetic fields.
We also thank the anonymous referee for valuable suggestions that improved
the quality of the paper.
DAZ thanks Pirinem School of Theoretical Physics for hospitality. 
The work of AST and AYP was supported by
the Russian Science Foundation grant 19-12-00133-P.
Scientific results reported in this paper are based on observations 
obtained with \xmm, en ESA science mission with instruments 
and contributions directly funded by ESA Member States and NASA.

\section*{Data Availability}

The \xmm\ data are available through the data 
archive \url{https://www.cosmos.esa.int/web/xmm-newton/xsa}.



\bibliographystyle{mnras}
\bibliography{ref} 



\appendix

\section{Pulsations detection significance and spin frequency uncertainty}
\label{a:simulations}

\begin{figure}
\begin{minipage}[h]{1.\linewidth}
\center{\includegraphics[width=1.0\linewidth,clip]{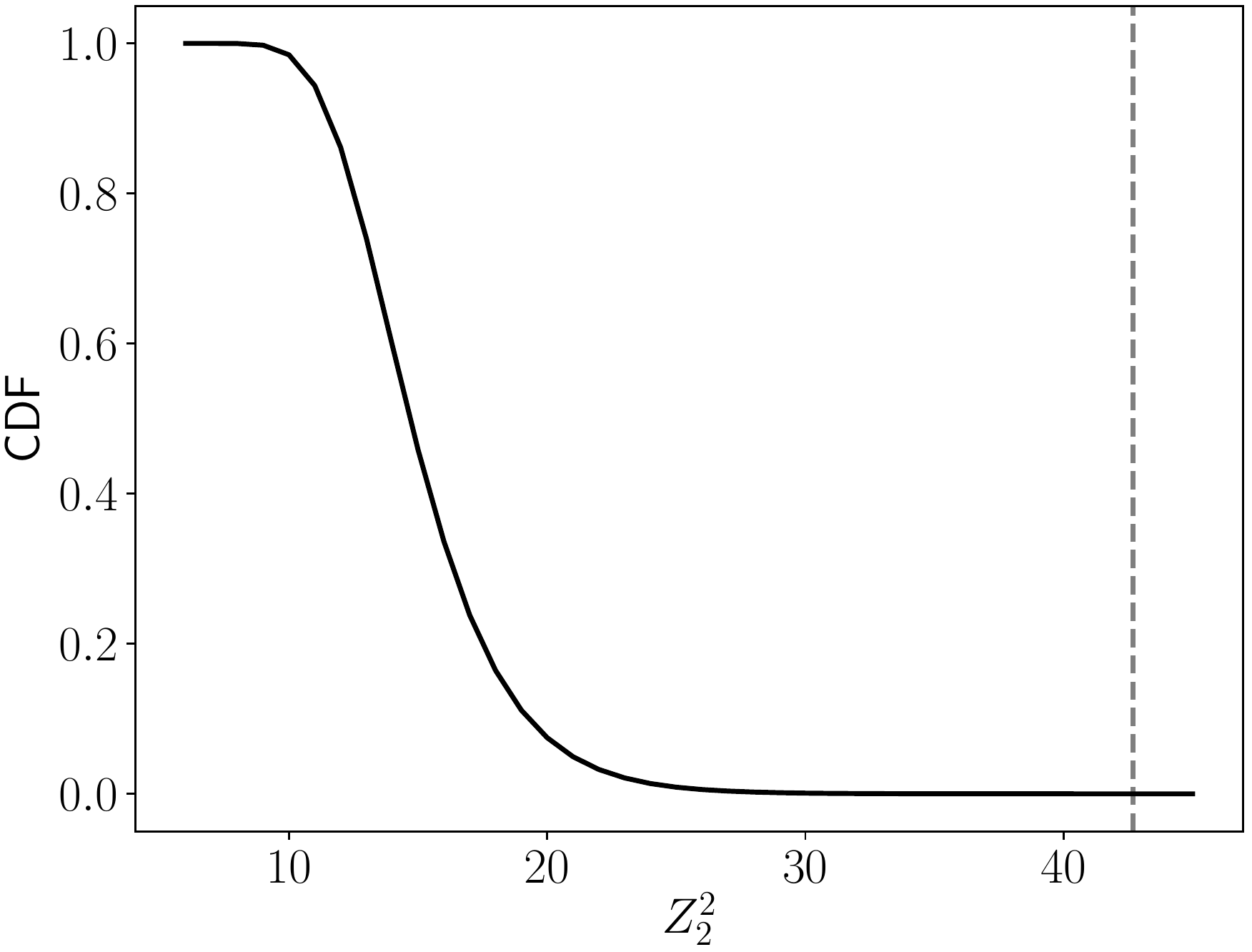}}
\end{minipage}
\caption{Cumulative distribution function of $Z_2^2$ for a pure noise signal with the mean 
count rate equal to the observed from \psr. The vertical dashed line corresponds to 
$Z_2^2 = 42.7$.
}
\label{fig:cdf}
\end{figure}

To estimate the confidence level of the detection of the periodic signal from 
\psr, we generated 1 million synthetic light curves with the length and the mean 
count rate equal 
to the ones observed from \psr, but consisting of pure Poisson noise without any 
periodic component. For each of the light curves, we performed $Z_2^2$-test 
using the same frequency window and the same number of trial frequencies as 
for $Z_2^2$-test on the observed light curve. The obtained highest $Z_2^2$ values 
were used to construct the cumulative distribution function (CDF) of $Z_2^2$ 
in the absence of the periodic signal (see Fig.~\ref{fig:cdf}). We found that 
the probability to get $Z_2^2 = 42.7$ from pure noise is about $2\times10^{-6}$, 
which corresponds to the detection confidence level of $\approx 4.7\sigma$.

\begin{figure}
\begin{minipage}[h]{1.\linewidth}
\center{\includegraphics[width=1.0\linewidth,clip]{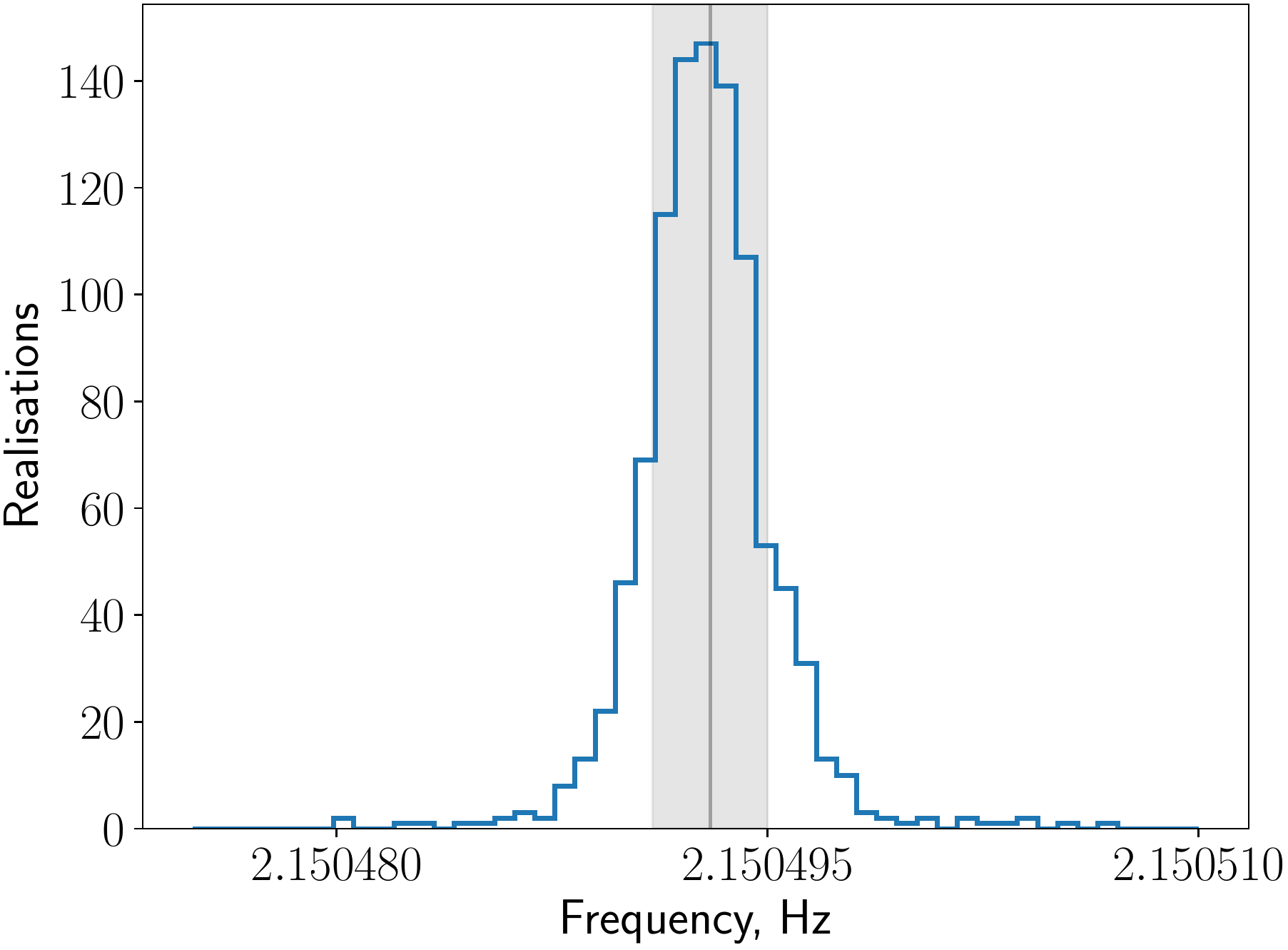}}
\end{minipage}
\caption{Distribution of the best frequencies measured for 1000 simulated event lists 
with the parameters fixed at the values obtained for \psr. The solid vertical line corresponds 
to the most probable value, while the shaded area indicates the 68 per cent credible interval
which was taken as the error of the frequency determination.
}
\label{fig:freqerr}
\end{figure}

The frequency uncertainty was estimated in a similar fashion. 
We fitted the pulse profile with the sum of the first two harmonics. We simulated 
1000 event lists of the periodic signal with the measured frequency, 
amplitudes and relative phases of two harmonics, keeping the mean count rate fixed 
and varying the number of photons and their times of arrival according to 
the Poisson statistics. 
For each event list, we found the frequency of the signal performing 
$Z_2^2$-test identical to the one applied to the real data. 
The nearly symmetrical 68 per cent confidence interval of 
the resulting frequency distribution (see Fig.~\ref{fig:freqerr}) was taken 
as the desired uncertainty.

\section{Atmosphere models}
\label{sec:atm}

\begin{figure}
\includegraphics[width=\columnwidth]{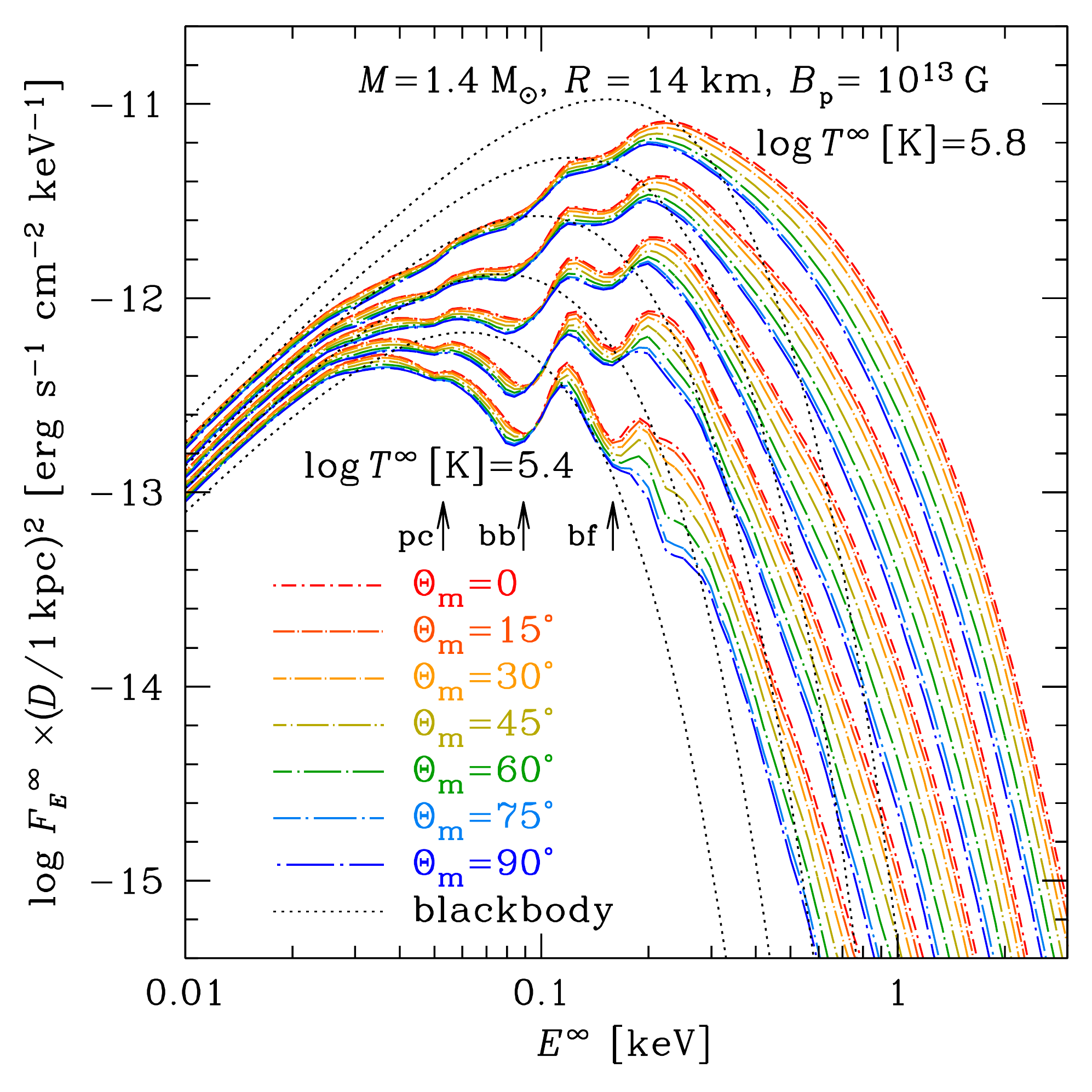}
\includegraphics[width=\columnwidth]{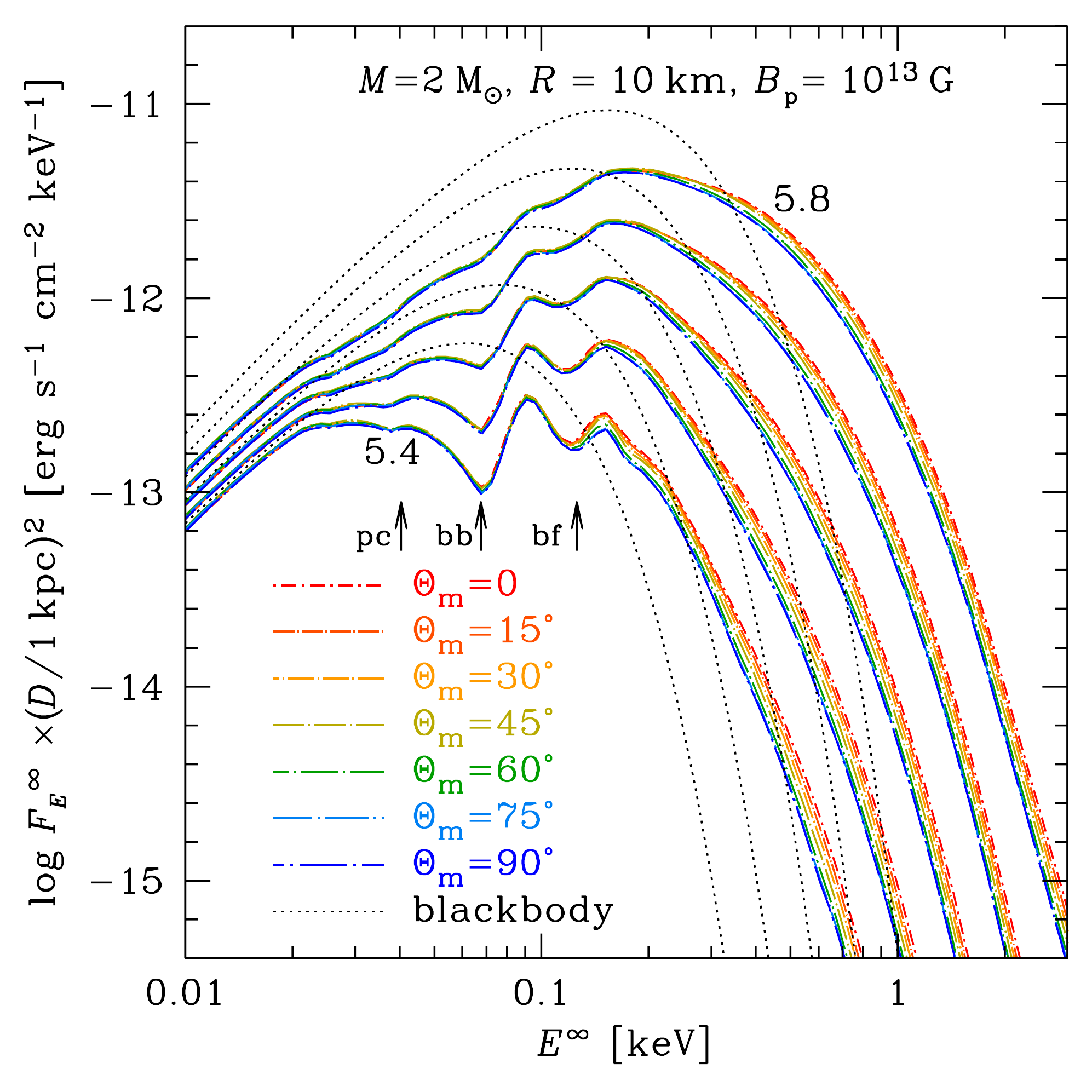}
\caption{Top panel: Thermal spectra of an NS model with $M=1.4\msun$, $R=14$~km and $B_\mathrm{p}=10^{13}$~G, 
as seen by a distant
observer, for $\log T_\mathrm{eff}^\infty\mbox{\,[K]}=5.4$, 5.5, 5.6,
5.7 and 5.8 (the five bunches of lines from bottom to top), and  seven
values of the angle between the magnetic axis and the line of sight,
$\Theta_\mathrm{m}$ (different line styles and colors according to the
legend). The arrows indicate approximate positions of the features due
to the proton cyclotron absorption (pc), the principal bound-bound
transition (bb) and the principal photoionization threshold (bf),
smeared by the field distribution over the surface and/or by atomic
motion effects. For comparison, the dotted lines show the blackbody
spectra for the same NS parameters and redshifted effective temperatures
$T^\infty$.
Bottom panel: The same as in the top panel but for an NS model with $M=2\msun$ and $R=10$~km.
}
\label{fig:spec_theor}
\end{figure}

The computation of the spectrum that can be measured by a distant
observer is patterned after \citet{Zyuzin_21}.  We construct the
integral spectrum by assembling local spectra at different patches on
the surface. We assume a dipolar magnetic field, modified by the effects
of General Relativity \citep{GinzburgOzernoy,PavlovZavlin00}. The
temperature distribution, which is associated with this magnetic field,
is calculated following \citet{Potekhin_03}. 

For every selected field strength at the magnetic pole $B_\mathrm{p}$
and selected NS mass and radius, the local radiative flux density was
computed at three magnetic latitudes, including the pole, for a set of
480 directions of the photon wave vector and for 150--200 photon
energies in the X-ray band, using an advanced version of the code
described in \citet*{SuleimanovPW09}. The fourth latitude is the
equator, which is too cold to allow construction of an atmosphere model
with the currently available opacities in strong magnetic fields
for the selected range of effective temperatures log$T^\infty$~[K] = 5.4--5.8. 
However, its contribution to the total flux is small, so we replace it by the
blackbody model (we have checked that using alternative models does not
lead to a noticeable change in the total spectrum). 

The code of \citet{SuleimanovPW09} has been modified to allow for
different angles $\theta_B$ between the magnetic field and the normal to
the surface. Hydrogen composition is considered, taking into account
incomplete ionization. The effects of the strong magnetic field and the
atomic thermal motion across the field on the plasma opacities are
treated following \citet{PotekhinChabrier03} with the improvements
described in \citet*{PotekhinCH14}. Polarization vectors and opacities
of normal electromagnetic modes are calculated as in
\citet{Potekhin_04}. Then flux values at arbitrary latitudes, energies
and directions are obtained by interpolation (or extrapolation,
whenever needed).

The monochromatic spectral flux density measured by a distant observer
is computed  by integrating the emission from different local patches
over the stellar surface for any selected angle $\Theta_\mathrm{m}$
between the magnetic dipole axis and the line of sight
(see Appendix~A of \citealt{Zyuzin_21} for details). 
In the axisymmetric model, the pulsar geometry is determined by the
angles $\alpha$ and $\zeta$ that the spin axis makes with the magnetic
axis and with the line of sight, respectively
\citep[e.g.,][]{PavlovZavlin00}. To produce phase-resolved spectra, it
is sufficient to calculate  $\cos\Theta_\mathrm{m} =
\sin\zeta\sin\alpha\cos\phi + \cos\alpha\cos\zeta$ for each rotation
phase $\phi$. Then the light curve and the phase-integrated spectrum are
given by integration of the phase-resolved spectrum over the energy or
over the phase $\phi$, respectively.

For isolated pulsars, the widely used
estimate of the magnetic field strength
is based on the expression
\begin{equation}
B \approx 3.2\times10^{19} \,C\,
 \sqrt{P\dot{P}}\textrm{~~G},
\label{PPdot} 
\end{equation}
where  $P$ is the period in seconds, $\dot{P}$ is the period time
derivative, and $C$ is a coefficient, which depends on stellar
parameters.  For the non-relativistic rotating magnetic dipole in vacuo
\citep{Deutsch55}, the magnetic field strength at the equator
$B_\mathrm{eq}$ is given by setting
\begin{equation}
C=R_{10}^{-3}\,(\sin\alpha)^{-1}\,\sqrt{I_{45}},
\label{Cdip}
\end{equation}
where
$R_{10}\equiv R/(10\mbox{~km})$ and $I_{45}$ is the moment of inertia in
units of $10^{45}$~g~cm$^2$. The latter depends on the EoS, but in most
plausible settings it can be estimated with an accuracy within 10 per cent by
the approximation of \citet{RavenhallPethick94}, which can be written in
the form
\begin{equation}
   I_{45} \approx 0.42(M/\msun)(R^\infty/10\mbox{~km})^2
\label{I45}
\end{equation}
(see \citealt{BejgerHaensel02} for a more general fitting formula). The
characteristic field $B_\mathrm{c}$ (Table~\ref{tab:pars}) is defined by
equation~(\ref{PPdot}) with $C=1$ \citep[e.g.,][]{atnf2005}. For the
likely values of $\alpha\gtrsim50^\circ$, $M\approx(1.6$--2.1)~\msun\ 
and $R\approx 11.7$--14.7~km, implied by the fitting results in
Table~\ref{tab:mcmc:bestfit}, equations (\ref{PPdot})--(\ref{I45})
give $B_\mathrm{eq}\sim(4$--$11)\times10^{12}$~G, which implies, for
the non-relativistic dipole field,
$B_\mathrm{p}\sim(0.8$--$2.2)\times10^{13}$~G. These values are
consistent with the values  $B_\mathrm{p}=10^{13}$~G and
$B_\mathrm{p}=2\times10^{13}$~G that we used to construct the atmosphere
models. We have also tried models with lower and higher field strengths,
but found that they do not provide a better fit.

A real pulsar differs from a rotating magnetic
dipole, because its magnetosphere is filled with plasma,
carrying electric charges and currents.
According to the results of numerical simulations of plasma
behavior in the pulsar magnetosphere \citep{Spitkovsky06}, the
equatorial magnetic field can be approximately
described by equation~(\ref{PPdot}) with 
\begin{equation}
   C\approx(0.8\pm0.1) R_{10}^{-3}\,(1+\sin^2\alpha)^{-1/2}
   \,\sqrt{I_{45}},
\label{CSpit}
\end{equation}
which gives estimates in the range of
$B_\mathrm{p}\sim(3$--$9)\times10^{12}$~G. Additional uncertainties
arise from the effects of General Relativity,  pulsar wind and
deviations from the pure dipole geometry (see \citealt{Petri19} for
discussion and references).

For each of the selected $B_\mathrm{p}$ values, we have considered $M$, $R$,
$T^\infty$, $\alpha$ and $\zeta$ as fitting parameters, using
interpolation and extrapolation based on the
computed spectra for $M=1.4\msun$ and 2.0\msun, $R=10$~km, 12~km and
14~km, $\log T^\infty\mbox{\,[K]}=5.4$, 5.5, 5.6, 5.7 and 5.8, and
various $\Theta_\mathrm{m}$. Examples
of the computed spectra are shown in the top and bottom panels of
Fig.~\ref{fig:spec_theor} for less and more compact NS models, respectively.


\bsp	
\label{lastpage} 

\end{document}